\documentclass[journal, 10pt, twocolumn, final]{IEEEtran}
\usepackage{epsfig}
\usepackage{amsmath,amssymb,amstext,amsthm}
\usepackage{MnSymbol}
\usepackage{latexsym,amssymb,amsmath,epsfig,subfigure,algorithm}
\usepackage{algorithmic}
\usepackage{epstopdf}
\usepackage{graphicx} 
\usepackage{ifthen}
\usepackage{color}
\usepackage{psfrag}
\usepackage{flushend}

\begin{document}
\ifCLASSOPTIONdraftcls
\title{Efficient Detectors for Telegram Splitting\\[-0.3cm] based Transmission in Low Power Wide Area \\[-0.3cm] Networks with Bursty Interference}
\else
\title{Efficient Detectors for Telegram Splitting\\ based Transmission in Low Power Wide Area \\ Networks with Bursty Interference}
\fi

\author{\IEEEauthorblockN{Steven Kisseleff,~\IEEEmembership{Member,~IEEE}, Jakob Kneissl, Gerd Kilian,~\IEEEmembership{Member,~IEEE}}, \\
and Wolfgang H. Gerstacker,~\IEEEmembership{Senior Member,~IEEE}
\thanks{
This paper has been presented in part at IEEE Global Communications Conference (Globecom) 2018 \cite{own}.

Steven Kisseleff is with the Interdisciplinary Centre for Security, Reliability and Trust (SnT), University of Luxembourg, Luxembourg, E-mail: steven.kisseleff@uni.lu.

J. Kneissl and G. Kilian are with the Radio Communication Systems Department, Fraunhofer Institute for Integrated Circuits (IIS), Erlangen, Germany, E-mail: \{jakob.kneissl,\ gerd.kilian\}@iis.fraunhofer.de.

Wolfgang H. Gerstacker is with the Institute for Digital Communications, Friedrich-Alexander University Erlangen-N\"urnberg (FAU), Erlangen, Germany, E-mail: wolfgang.gerstacker@fau.de.
} 
}
\maketitle
\thispagestyle{empty}
\begin{abstract}
Low Power Wide Area (LPWA) networks are known to be highly vulnerable to external in-band interference in terms of packet collisions which may substantially degrade the system performance. In order to enhance the performance in such cases, the telegram splitting (TS) method has been proposed recently. This approach exploits the typical burstiness of the interference via forward error correction (FEC) and offers a substantial performance improvement compared to other methods for packet transmissions in LPWA networks. While it has been already demonstrated that the TS method benefits from knowledge on the current interference state at the receiver side, corresponding practical receiver algorithms of high performance are still missing.
The modeling of the bursty interference via Markov chains leads to the optimal detector in terms of a-posteriori symbol error probability. However, this solution requires a high computational complexity, assumes  an a-priori knowledge on the interference characteristics and lacks flexibility. We propose a further developed scheme with increased flexibility and introduce an approach to reduce its complexity while maintaining a close-to-optimum performance. In particular, the proposed low-complexity solution substantially outperforms existing practical methods in terms of packet error rate and therefore is highly beneficial for practical LPWA network scenarios.
\end{abstract}

\begin{IEEEkeywords}
Low power wide area networks, interference, Markov chains, maximum a-posteriori probability detection.
\end{IEEEkeywords}
\section{Introduction}
\label{sec_1}
\subsection{Background}
\IEEEPARstart{M}{assive} telemetry systems are considered an important part of the upcoming Internet-of-Things (IoT) \cite{7815384}. Such systems may consist of thousands of small sensor/metering devices, which are battery-powered and transmit the sensed information over a large distance such as 10 km or more for collection at a central gateway. Hence, they are typical examples of so-called Low Power Wide Area Networks (LPWANs). An important challenge for such systems is a large amount of interference created by the network participants, which may lead to packet loss due to collisions and correspondingly to frequent packet retransmissions. One of the most common LPWAN system concepts is LoRaWAN \cite{8030482}, which utilizes ALOHA protocols for packet retransmissions and channel hopping, cf. \cite{7721743}, and is supposed to provide connectivity to thousands of sensing devices in a limited area. In this context, strong interference may affect many packets and limit the performance of LoRaWAN in particular and LPWANs in general.
In addition, LPWANs are supposed to coexist with other wireless systems, as explained in \cite{7132717} and \cite{8422801}. 
Many LPWAN systems operate in the license exempt frequency bands. For example, in Europe the short range devices (SRD) bands from 863 MHz to 870 MHz and around 2.4 GHz are most relevant. The SRD bands are utilized by a variety of different applications such as radio frequency identification (RFID) systems, wireless microphones or other communication standards based on IEEE 802.15.4, cf. \cite{8422801}. It can be expected that the channel load in the SRD bands will continue to grow significantly and therefore the external interference to LPWAN systems resulting in a high probability of collisions. Hence, in this paper we focus on LPWAN transmissions which are typically ultra-narrow band, interfered by short wide band transmissions of other systems. 
%

The harmful influence of packet collisions can be reduced using forward error correction (FEC) coding. However, the choice of the FEC code implies a trade-off between the packet error rate (PER) and the energy consumption, since in general more redundant symbols have to be transmitted with increasing strength of the selected channel code. In addition, FEC coding may not be useful in the presence of a vast amount of interference as mentioned in \cite{7575656}, especially if all symbols of the packet are overlapped by an unknown number of interferers with variable signal power. In \cite{6525243}, the telegram splitting (TS) method has been introduced\footnote{Telegram splitting has been standardized for ultra-narrow band communication systems according to ETSI TS 103357.}, where the bursty behavior of the interference is exploited which stems from the fact that also the sensing information acquired by other systems may be transmitted discontinuously. Hence, there might be interference scenarios which are more suitable for the application of FEC coding. In order to increase the probability that at least a part of the packet would be received with a sufficiently low amount of interference, the packet (called telegram in \cite{6525243}) is split in multiple sub-packets, which are transmitted independently with a certain time and frequency spacing between them. At the receiver, the sub-packets are reassembled and an attempt is made to correct the errors within the more damaged sub-packets using FEC. In \cite{6525243} and \cite{6999938}, the benefit of this method compared to the traditional packet-by-packet transmission without TS in terms of reliability of the signal detection has been demonstrated. Furthermore, an optimization of the number of sub-packets and the coding rate has revealed that the PER can be substantially reduced by increasing the number of sub-packets, i.e., splitting the telegram into many parts. In this paper, we focus on the TS method for LPWAN transmission.

In order to reach a sufficiently high signal quality, some of the LPWANs that utilize relatively high symbol rates, such as Narrow-Band (NB) IoT, may employ tens or hundreds of packet repetitions in order to reach a sufficiently high coupling loss. However, LPWAN schemes such as LoRaWAN, Sigfox and ETSI TS 103357 follow a different approach and employ a reduced symbol rate such that it is possible to reach a high coupling loss with few or no packet repetitions. In fact, Sigfox utilizes three packet repetitions with ultra-narrow band transmissions while LoRa uses a spread-spectrum technique. ETSI TS 103357 employs telegram splitting as considered in our paper and a very strong FEC code with low code rate of $1/3$, and does not need to apply any explicit repetitions for a good performance.
\vspace*{-2mm}
\subsection{Previous works}
For the TS method, the knowledge of the interference variance in the presence of bursty interference is important in order to distinguish between the corrupted and "clean" sub-packets at the receiver. For this, the interference states have been modeled using Markov chains in \cite{7575656}. However, only the presence or absence of interference is predicted based on the state estimates of the respective Markov chain. The corrupted symbols are excluded (erased) from the decoding process. This solution is suboptimal due to the complete erasure of the corrupted symbols, which may still carry a certain amount of useful information.

A related problem of symbol detection in the presence of bursty noise is well known in the context of powerline communications (cf. \cite{6547827}, \cite{990732}), and the methods which have been proposed to tackle the problem in this field are discussed in the following.

In powerline communications, abrupt discharging effects within various segments of the powerline can heavily affect the signal transmission. Such effects are very difficult to model due to the intricate structure of the powerline system, such that machine learning approaches (as in \cite{6547827}) and methods based on Markov chains (as in \cite{990732}) seem promising. Also, it has been observed that the noise in powerline communications may exhibit a certain structure which can be characterized and exploited for symbol detection. More general scenarios for symbol detection in the presence of bursty noise have been studied in \cite{fertonani2009reliable} and \cite{5504595}. For this, a Bernoulli-Gaussian channel has been assumed. Transition probabilities between the "good" and "bad" channel states have been determined, and Markov chains have been established for channel description. Here, a "good" state implies that a reliable symbol detection is possible in principle, whereas for a bad state, no successful detection is possible due to a very large noise variance. Hence, similar to \cite{7575656}, the corrupted symbols corresponding to a "bad" state are discarded.

Unlike in \cite{7575656} and \cite{5504595}, a novel maximum a-posteriori (MAP) symbol detection scheme for bursty external interference has been proposed in \cite{own}, where it has been assumed that all realizations of the external interference belong to a single communication system with known signal characteristics, i.e., signal variance, packet length and arrival probability (similar to \cite{6999938}). The method of \cite{own} exploits these properties of the external interference which are assumed to be known at the receiver, using a specifically designed Markov chain in order to maximize the reliability of symbol detection. However, in general, the interference characteristics are not perfectly known at the receiver. Furthermore, the external interferers may belong to multiple communication systems or different modes of a system with individual duty cycles and packet lengths, cf. \cite{8422801}, \cite{6824847}. Hence, from a practical point of view, a detection scheme should be able to cope with multiple classes of interferers and also require only a limited a priori knowledge of the interference characteristics. Another disadvantage of the method proposed in \cite{own} is related to its very high computational complexity, which restricts the application of the method to specific scenarios.
\vspace*{-2mm}
\subsection{Contribution}
In this paper, we extend the optimal MAP detection method for bursty external interference described in \cite{own} such that multiple classes of interferers are admissible. The resulting optimal detector shows a  computational complexity which is too high for practical applications in most cases. Nevertheless, the method can be used in order to determine an upper bound for the performance of any suboptimum detector. In order to reduce the complexity of the optimum scheme, we propose some modifications to the original algorithm which do not lead to a significant performance degradation. However, the resulting complexity reduction is still not sufficient for scenarios with many classes of interferers.

Furthermore, we propose a new method which shows a close-to-optimal performance (i.e., close to the performance of the optimal MAP detector) under very low and adjustable complexity. Moreover, this method does not require any prior knowledge of the structure of the interference at the receiver, but is solely based on the long-term observation of the received signal in an initial phase. Correspondingly, this new method can be employed for an arbitrary number of interference classes and shows a high flexibility in general. Its adjustable complexity and high performance render it well suitable for practical applications.

Our contributions are summarized as follows:
\begin{itemize}
\item Development of the optimal MAP detection algorithm for LPWAN transmission in the presence of bursty external interference and multiple interference classes;
\item Design of a modified version of the optimal MAP detector with reduced complexity and only small degradation in performance;
\item Design of a close-to-optimum adaptive low complexity detector with adjustable complexity, requiring no prior information about the interference scenario.
\end{itemize}

Although the proposed detection algorithms were originally intended to be employed in conjunction with the TS method as the best approach for a LPWAN transmission, they can be easily incorporated in existing LPWAN systems which do not utilize TS as well, e.g. the LoRaWAN system. In particular, since the proposed low complexity  detector does not require any prior knowledge on the external interference and is solely based on a long-term observation of the received signal, it can be adjusted during the initialization phase of LoRaWAN \cite{7721743}.

This paper is organized as follows. In Section \ref{sec_2}, the underlying system model is described. Several MAP based detection schemes for bursty external interference based on interference modeling are proposed in Section \ref{sec_3}. We start with the optimal full-state MAP detector for a single interference class and extend it to a multiclass MAP detector and a reduced-state MAP detector, respectively. Then, a practical low complexity algorithm is proposed. The performance of the proposed schemes is evaluated via simulations and compared with relevant previously proposed schemes in Section \ref{sec_4}. Subsequently, conclusions are given in Section \ref{sec_5}.
\section{System Model}
\label{sec_2}
In this work, we assume a system model that is similar to the system model described in \cite{6999938}. We consider a single LPWAN uplink transmission from the node devices to the base station impaired by $N$ interfering signals. The structure of the system model is shown in Fig.\ \ref{chain:fig}. The transmitter encodes and modulates the data of a telegram before splitting it in multiple sub-packets and sending them to the receiver. The received signal comprises the sub-packets of interest, noise and multiple interferers.
\ifCLASSOPTIONdraftcls
\begin{figure}
\centering
\psfrag{Encoder}{\begin{small}Encoder\end{small}}
\psfrag{Shit}{\begin{small}Telegram\end{small}}
\psfrag{split}{\begin{small}splitting\end{small}}
\psfrag{ass}{\begin{small}assembling\end{small}}
\psfrag{IF1}{\begin{small}Interferer 1\end{small}}
\psfrag{IFN}{\begin{small}Interferer $N$\end{small}}
\psfrag{n}{\begin{small}noise\end{small}}
\psfrag{LLR}{\begin{small}LLR\end{small}}
\psfrag{gfgserj}{\begin{small}calculation\end{small}}
\psfrag{Decoder}{\begin{small}Decoder\end{small}}
\psfrag{Tx}{\begin{normalsize}Transmitter\end{normalsize}}
\psfrag{Rx}{\begin{normalsize}Receiver\end{normalsize}}
\psfrag{Data}{\begin{normalsize}Data\end{normalsize}}
\psfrag{output}{\begin{normalsize}Decoded data\end{normalsize}}
\includegraphics[width=0.8\textwidth]{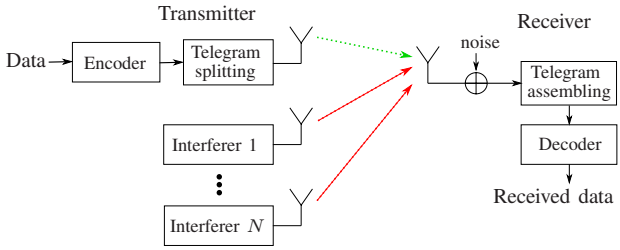}
\caption{System model including the transmitter with TS, $N$ interferers and the receiver.}
\label{chain:fig}
\vspace*{-2mm}
\end{figure}
\else
\begin{figure}
\psfrag{Encoder}{\begin{scriptsize}\hspace*{-1mm}Encoder\end{scriptsize}}
\psfrag{Shit}{\begin{scriptsize}\hspace*{-1mm}Telegram\end{scriptsize}}
\psfrag{split}{\begin{scriptsize}\hspace*{-1mm}splitting\end{scriptsize}}
\psfrag{ass}{\begin{scriptsize}\hspace*{-1mm}assembling\end{scriptsize}}
\psfrag{IF1}{\begin{scriptsize}\hspace*{-1mm}Interferer 1\end{scriptsize}}
\psfrag{IFN}{\begin{scriptsize}\hspace*{-1mm}Interferer $N$\end{scriptsize}}
\psfrag{n}{\begin{scriptsize}noise\end{scriptsize}}
\psfrag{LLR}{\begin{small}LLR\end{small}}
\psfrag{gfgserj}{\begin{scriptsize}calculation\end{scriptsize}}
\psfrag{Decoder}{\begin{scriptsize}Decoder\end{scriptsize}}
\psfrag{Tx}{\begin{footnotesize}Transmitter\end{footnotesize}}
\psfrag{Rx}{\begin{footnotesize}Receiver\end{footnotesize}}
\psfrag{Data}{\begin{footnotesize}Data\end{footnotesize}}
\psfrag{output}{\begin{footnotesize}\hspace*{-3mm}Received data\end{footnotesize}}
\includegraphics[width=0.45\textwidth]{kisse1.eps}
\caption{System model including the transmitter with TS, $N$ interferers and the receiver.}
\label{chain:fig}
\vspace*{-2mm}
\end{figure}
\fi

For the composition of the transmit signal, we assume a sequence of $k$ information bits which are encoded using an $(n, k)$-convolutional code
with code rate $R_c=k/n$ and pseudo-randomly interleaved. Furthermore, the encoded bits are mapped onto binary phase-shift keying (BPSK) symbols, such that a sequence of symbols $c[l]\in \{-1, +1\},\: 0\leq l<n$ results. According to the TS principle, the total symbol packet of length $n$ is split into multiple blocks (sub-packets) with $L_S$ symbols each\footnote{The number of information bits $k$ and the code rate $R_c$ are selected such that $n$ equals a multiple of $L_S$. Furthermore, $L_S$ is an even number.}. In addition, for synchronization purposes, a training sequence $r[\nu],\: 0\leq \nu<L_{\mathrm{tr}}$ with $L_{\mathrm{tr}}$ symbols is inserted in the middle of each sub-packet. Hence, the total length of the resulting sub-packet is equal to $L_\mathrm{tot}=L_S+L_{\mathrm{tr}}$. The sequence of symbols within a sub-packet with index $i$, $i \in \{ 1, \, 2, \, \ldots, \, n/L_S \}$ is denoted as $x_i[m]$ and given by
\ifCLASSOPTIONdraftcls
\begin{equation}
\label{eq_0}
x_i[m]=\left\{\begin{array}{ll}
c[m+L_S\cdot (i-1)], & 0\leq m< L_S/2\\
r[m-L_S/2], & L_S/2\leq m <L_S/2+L_{\mathrm{tr}}\\
c[m+L_S\cdot i-L_{\mathrm{tot}}], & L_S/2+L_{\mathrm{tr}}\leq m<L_\mathrm{tot}.
\end{array}\right.
\end{equation}
\else
\begin{equation}
\label{eq_0}
x_i[m]\hspace*{-1mm}=\hspace*{-1mm}\left\{\begin{array}{ll}
\hspace*{-1.5mm}c[m+L_S\cdot (i-1)], & \hspace*{-3mm}0\leq \hspace*{-0.5mm}m\hspace*{-0.5mm}< L_S/2\\
\hspace*{-1.5mm}r[m-L_S/2], & \hspace*{-3mm}L_S/2\leq \hspace*{-0.5mm}m\hspace*{-0.5mm} <L_S/2+L_{\mathrm{tr}}\\
\hspace*{-1.5mm}c[m+L_S\cdot i-L_{\mathrm{tot}}], & \hspace*{-3mm}L_S/2+L_{\mathrm{tr}}\leq \hspace*{-0.5mm}m\hspace*{-0.5mm}<L_\mathrm{tot}.
\end{array}\right.
\end{equation}
\fi
The resulting baseband transmit signal $d_i(t)$ for sub-packet $i$ is given by
\begin{equation}
\label{eq_1}
d_i(t)=\sum_{m=0}^{L_\mathrm{tot}-1}x_i[m]\:g(t-m\cdot T),
\end{equation}
where $g(t)$ and $T$ are a square-root Nyquist transmit pulse and the symbol interval, respectively. For the signal transmission, we consider an additive white Gaussian noise (AWGN) channel with noise variance $\sigma_N^2$ since intersymbol interference effects can be neglected in LPWAN transmission because of a relatively low symbol rate $1/T$. The noise variance is assumed to remain constant during the transmission of the whole telegram which is justified by the fact that communication channels in LPWA networks are typically static. Furthermore, a matched filter with impulse response $g^*(-t)$ is applied at the receiver. The signal at the output of the matched filter is sampled at time instants $m \, T$ resulting in a sequence $y_i[m]$. According to the employed AWGN channel model and the square-root Nyquist property of $g(t)$,
\begin{equation}
y_i[m]=x_i[m]+n_i[m]+z_i[m]
\end{equation}
holds, where $n_i[m]$ is white discrete-time noise and $z_i[m]$ denotes the total external interference. 
For the detection and decoding, the training symbols are removed and the remaining data symbols from all sub-packets are deinterleaved. The decoding is done using a Viterbi algorithm with soft input obtained from the receive sequences $y_i[m],\: i \in \{ 1, \, 2, \ldots, \, n/L_S\}$ via a specifically designed detector. In this context, the optimal detector would determine the soft input in a way that it corresponds to the a-posteriori probabilities of the symbols. Such detector is referred to as MAP symbol detector \cite{proakis}.

The bursty external interference is modeled via a Poisson arrival process with a known arrival probability $p_a$ and possesses a constant duration of $L_I$ symbols for each occurrence. More specifically, the Poisson process governs the number of interference arrivals per symbol interval, and the arrival probability within one symbol interval is given by
\begin{equation}
\label{eq_2}
p_a=1-\exp(-G/L_I),
\end{equation}
where $G$ is the interference load, i.e., the average fraction of time for which the channel is occupied by interference normalized by the symbol duration\footnote{If the interference results e.g. from an external LoRaWAN system, the interference load corresponds to the accumulated duty cycle, where the duty cycle per node is $1\%$ \cite{8030482}. Depending on the number of nodes in proximity of the LPWAN receiver of interest, typical values of $G$ range between 0.1 and 0.6, since LoRaWAN supports up to 3000 devices on multiple parallel channels \cite{8030482}.}. According to \cite{markov}, such behavior can be modeled by a Markov chain. Although the interfering signals are symbols from discrete symbol constellations, the choice of these constellations is up to the respective interfering communication system and therefore unknown to the considered LPWAN receiver. Taking into account different symbol intervals and the absence of synchronization between the interfering communication system and the LPWAN system of interest, intersymbol interference between the symbols of the interfering signal arises after the sampling.
Therefore, the pdf of the total interfering signal can be well approximated by a normal distribution according to the central limit theorem. The approximation of $z_i[m]$ by discrete-time white Gaussian noise is even more accurate in case of a significantly higher bandwidth of the interfering system compared to the LPWAN system of interest. Hence, we assume that the sampled bursty interference follows a normal distribution with constant variance  $\sigma_I^2$. For more details on the modeling of the interference from wide-band SRD systems to LPWAN systems, we refer to \cite{8422801}.

For simplicity of modeling, each transmitter of the interfering system can start its transmission only exactly at the beginning of the respective symbol interval. This is a mild assumption which only removes boundary effects with respect to the interference. Here, each interference packet would overlap exactly $L_I$ symbols of the data packet. In addition, according to the Poisson process assumption, the simultaneous arrival of two or more interferers in the same symbol interval is very unlikely for practical system parameters. Hence, the maximum number of active interferers that can be observed within any symbol interval is $L_I$, and the maximum assumed interference variance is $L_I\sigma_I^2$ as can be seen in Fig. \ref{trellis:fig}.
\ifCLASSOPTIONdraftcls
\begin{figure}
\centering
\includegraphics[width=0.7\textwidth]{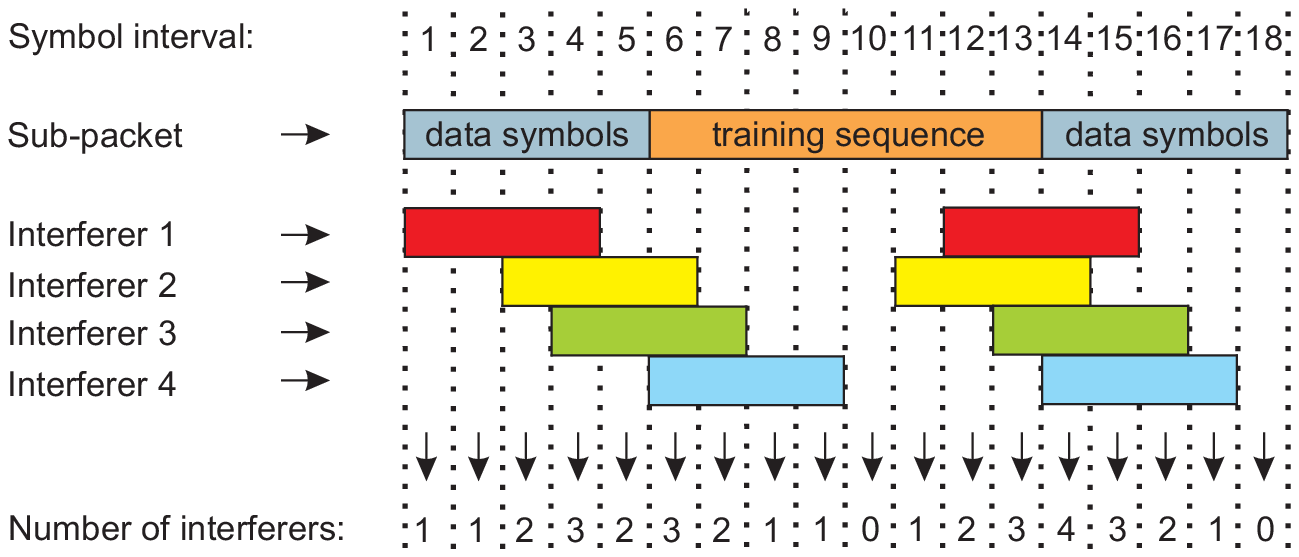}
\caption{Example of interference arrival during a sub-packet transmission with $L_S=10$, $L_{\mathrm{tr}}=8$, and $L_I=4$.}
\label{trellis:fig}
\vspace*{-2mm}
\end{figure}
\else
\begin{figure}
\includegraphics[width=0.49\textwidth]{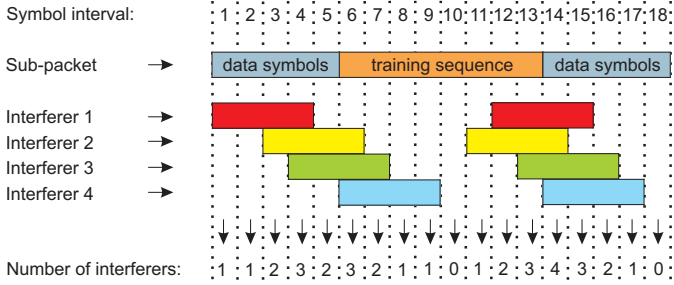}
\caption{Example of interference arrival during a sub-packet transmission with $L_S=10$, $L_{\mathrm{tr}}=8$, and $L_I=4$.}
\label{trellis:fig}
\vspace*{-2mm}
\end{figure}
\fi
Obviously, the total received interference is not memoryless if multiple consecutive symbol intervals are overlapped by the same interferer. However, the memory is reflected in a correlation of the interference variance for different symbol intervals instead of a correlation of the neighboring interference samples as it is typical for traditional transmission channels with memory (cf. \cite{proakis}). Nevertheless, such a dependency may carry enough information in order to enable some estimation of the interference variance and the reliability of the detected symbols. The reliability of signal detection can be expressed in terms of log-likelihood ratios (LLRs), which are typically used as soft input for the Viterbi algorithm for FEC. In the following, a novel detection scheme is proposed which determines the LLRs in an optimal way via a MAP symbol detection approach.
\section{Markov chain based detection}
\label{sec_3}
In this section, a Markov chain based detector is proposed for the computation of accurate a-posteriori symbol probabilities accounting for the nature of the external interference which are utilized for forming soft input to the Viterbi algorithm for a more efficient decoding. At first, a Markov chain model is developed for the interference. Then, the calculation of a-posteriori symbol probabilities is described. Furthermore, it turns out that the detection performance can be substantially improved if the knowledge of the training sequence at the receiver is exploited for detection, due to the memory of the interference. Hence, a correct incorporation of the training symbols into the a-posteriori symbol probability calculation is discussed as well.
\subsection{Markov chain model}
\label{sec_3_1}
In order to create a Markov chain model that would describe the arrival of interference blocks, a suitable state definition for the Markov chain is required. At first, we define an "interferer state" as the remaining number of symbols to be overlapped by that interferer including the current symbol. As an example, once an interferer of length four ($L_I=4$) arrives, the remaining number of symbols to be overlapped by this interferer will also be four. Obviously, in the next symbol interval, the state number of this interferer would reduce by one and becomes three, see Fig. \ref{states:fig}.
\ifCLASSOPTIONdraftcls
\begin{figure}
\centering
\includegraphics[width=0.7\textwidth]{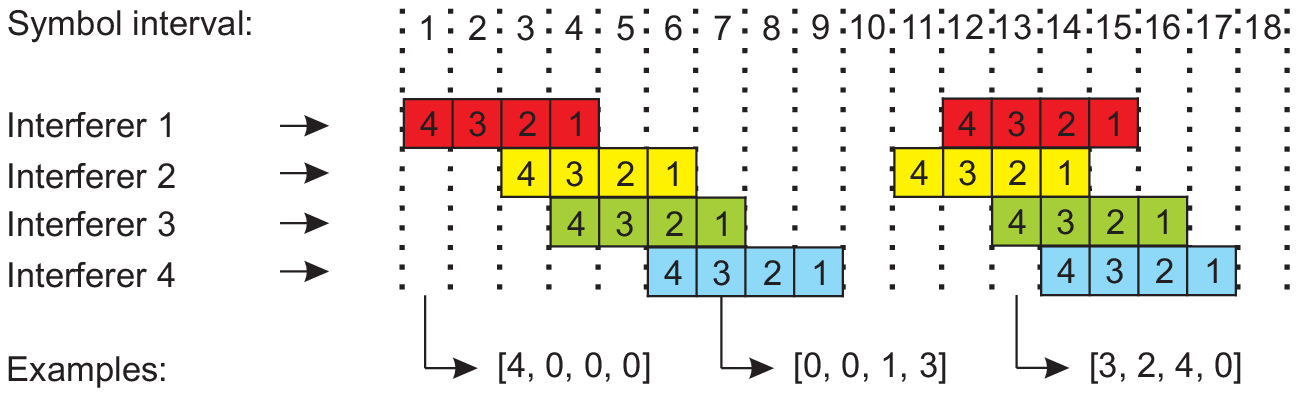}
\caption{Definition of interferer states.}
\label{states:fig}
\vspace*{-2mm}
\end{figure}
\else
\begin{figure}
\includegraphics[width=0.49\textwidth]{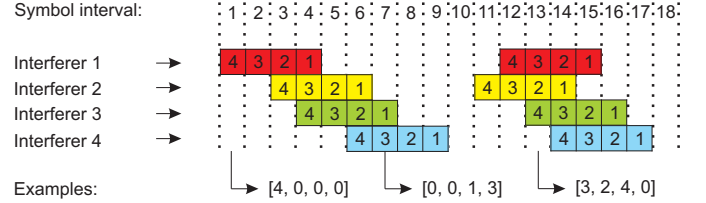}
\caption{Definition of interferer states.}
\label{states:fig}
\vspace*{-2mm}
\end{figure}
\fi
In absence of the interferer, the respective interferer state is set to '0'. Correspondingly, each total state $s$ of the Markov chain can be viewed as a member of a set $\mathcal{S}$, where the set members are vectors describing the states of all active interferers. For example, $\left[4, 0, 0, 0\right]$, $\left[0, 0, 1, 3\right]$, and $\left[3, 2, 4, 0\right]$ are possible total states in case of $L_I=4$, see Fig. \ref{states:fig}.

In the following, we determine the number of states needed in order to capture all combinations of substates of active interferers. For the calculation of the total number of states, we first assume $l$ active interferers, such that $L_I\cdot(L_I-1)\cdot(L_I-2)\ldots(L_I-l+1)=\frac{\left(L_I\right)!}{\left(L_I-l\right)!}=\binom{L_I}{l}\left(l\right)!$ is the number of combinations of active interferers, accounting for the interferer positions and the fact that each non-zero interferer substate number can arise only once in the total state due to the properties of the Poisson arrival process.
Within a vector of length $L_I$, each combination can be assigned to one of the $\binom{L_I}{l}$ sets of $l$ positions. Hence, the number of states with $l$ active interferers is $\binom{L_I}{l}^2\left(l\right)!$ and the total number of states with any number of active interferers is $\sum_{l=0}^{L_I}\binom{L_I}{l}^2\left(l\right)!$. 
Here, the order of interferer substates is taken into account, such that the two total states $\left[3, 0, 2, 0\right]$ and $\left[0, 3, 0, 2\right]$ are distinguished. Through this, the history of all (maximum $L_I$) individual interferers can be preserved. On the other hand, the mentioned two states represent the same number of active interferers, the same overall history, and the same current channel condition, since the variances of all interferers are assumed to be equal. Hence, the number of states can be reduced by removing states which are redundant in such sense from the state list. Subsequently, we sort the interferer substates of the remaining states according to the number of symbols still to be overlapped by the respective interferers in descending order. For the example above, both states would be represented by a single state $\left[3, 2, 0, 0\right]$ without any loss of information. Accordingly, the number of states in the Markov chain and set $\mathcal{S}$ reduces to 
$\sum_{l=0}^{L_I}\binom{L_I}{l}=2^{L_I}$ which corresponds to an enormous complexity reduction compared to the previous state definition. The number of states of the Markov chain model versus $L_I$ is depicted in Fig. \ref{Nstates}.
\ifCLASSOPTIONdraftcls
\begin{figure}
\centering
\psfrag{without}{\begin{small}without state sorting\end{small}}
\psfrag{with}{\begin{small}with state sorting\end{small}}
\psfrag{Number}{\begin{normalsize}Number of states\end{normalsize}}
\psfrag{length}{\begin{normalsize}Interferer length $L_I$\end{normalsize}}
\includegraphics[width=0.63\textwidth]{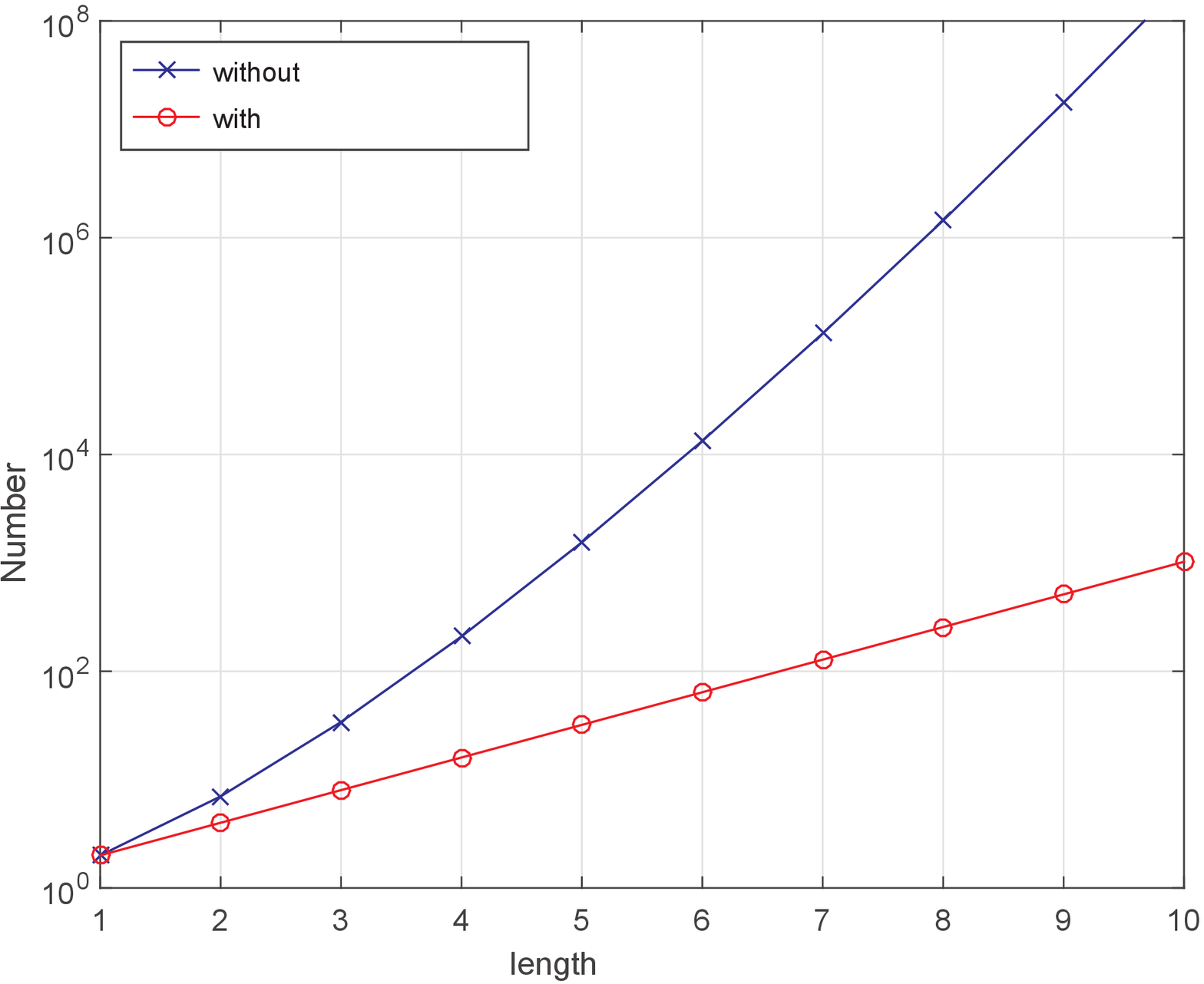}
\caption{Required number of states for modeling of the interference Markov process.}
\label{Nstates}
\vspace*{-2mm}
\end{figure}
\else
\begin{figure}
\psfrag{without}{\begin{scriptsize}without state sorting\end{scriptsize}}
\psfrag{with}{\begin{scriptsize}with state sorting\end{scriptsize}}
\psfrag{Number}{\begin{footnotesize}Number of states\end{footnotesize}}
\psfrag{length}{\begin{footnotesize}Interferer length $L_I$\end{footnotesize}}
\includegraphics[width=0.46\textwidth]{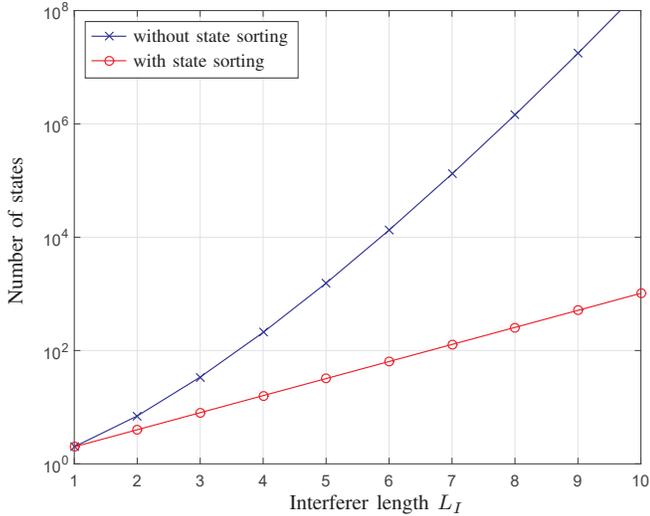}
\caption{Required number of states for modeling of the interference Markov process.}
\label{Nstates}
\vspace*{-2mm}
\end{figure}
\fi
For $L_I=6$, the Markov chain becomes too large ($>$ 10000 states) for a practical implementation, if no state sorting is applied. With state sorting, the number of states is 64 in this case which is much lower. For $L_I=2$, the number of states with and without sorting equals 4 and 7, respectively. Hence, the complexity can be reduced by at least 42$\%$ for $L_I \ge 2$.

So far, only the interferer states have been modeled. However, for the calculation of the a-posteriori symbol probabilities, we need to distinguish between different states of the useful signal as well. Assuming a BPSK modulation, we consider two separate Markov chains for the interferer states and data symbols $+1$ and $-1$, respectively. These two Markov chains are then combined into one (product-) chain. Correspondingly, we define the states of the product-chain as a concatenation of the interferer states and the current BPSK symbol, resulting in states such as $\left[3, 2, 0, 0, +1\right]$ or $\left[4, 3, 2, 0, -1\right]$. The corresponding sets of states pertaining to $+1$ and $-1$ are denoted as $\mathcal{S}_+$ and $\mathcal{S}_-$, respectively. 
The resulting Markov product-chain for $L_I=2$ is shown in Fig. \ref{Markov}. 
\ifCLASSOPTIONdraftcls
\begin{figure}
\centering
\psfrag{00+}{\begin{normalsize} $[0, 0, +1]$\end{normalsize}}
\psfrag{10+}{\begin{normalsize} $[1, 0, +1]$\end{normalsize}}
\psfrag{20+}{\begin{normalsize} $[2, 0, +1]$\end{normalsize}}
\psfrag{21+}{\begin{normalsize} $[2, 1, +1]$\end{normalsize}}
\psfrag{00-}{\begin{normalsize} $[0, 0, -1]$\end{normalsize}}
\psfrag{10-}{\begin{normalsize} $[1, 0, -1]$\end{normalsize}}
\psfrag{20-}{\begin{normalsize} $[2, 0, -1]$\end{normalsize}}
\psfrag{21-}{\begin{normalsize} $[2, 1, -1]$\end{normalsize}}
\psfrag{p}{\begin{small}$p$\end{small}}
\psfrag{n}{\begin{small}$q$\end{small}}
\includegraphics[width=0.8\textwidth]{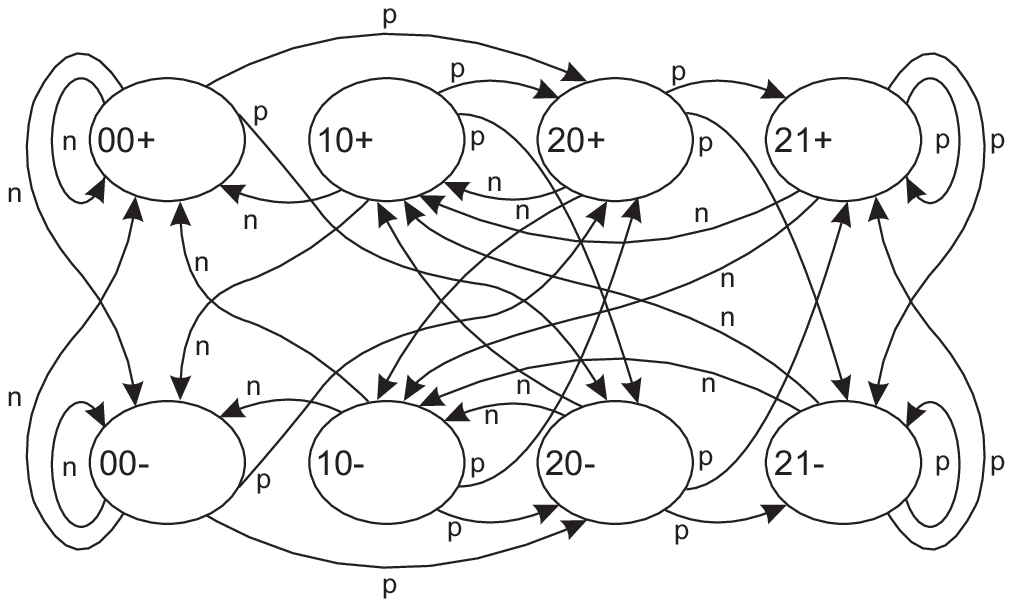}
\caption{A Markov product-chain describing the interferer states for $L_I=2$. Here, $p=\frac{1}{2}p_a$ and $q=\frac{1}{2}(1-p_a)$ holds.}
\label{Markov}
\vspace*{-2mm}
\end{figure}
\else
\begin{figure}
\psfrag{00+}{\begin{footnotesize}$[0, 0, +1]$\end{footnotesize}}
\psfrag{10+}{\begin{footnotesize}$[1, 0, +1]$\end{footnotesize}}
\psfrag{20+}{\begin{footnotesize}$[2, 0, +1]$\end{footnotesize}}
\psfrag{21+}{\begin{footnotesize}$[2, 1, +1]$\end{footnotesize}}
\psfrag{00-}{\begin{footnotesize}$[0, 0, -1]$\end{footnotesize}}
\psfrag{10-}{\begin{footnotesize}$[1, 0, -1]$\end{footnotesize}}
\psfrag{20-}{\begin{footnotesize}$[2, 0, -1]$\end{footnotesize}}
\psfrag{21-}{\begin{footnotesize}$[2, 1, -1]$\end{footnotesize}}
\psfrag{p}{\begin{scriptsize}$p$\end{scriptsize}}
\psfrag{n}{\begin{scriptsize}$q$\end{scriptsize}}
\includegraphics[width=0.49\textwidth]{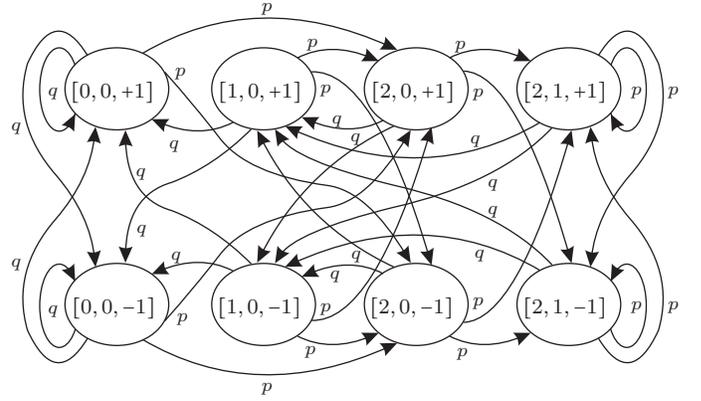}
\caption{A Markov product-chain describing the interferer states for $L_I=2$. Here, $p=\frac{1}{2}p_a$ and $q=\frac{1}{2}(1-p_a)$ holds.}
\label{Markov}
\vspace*{-2mm}
\end{figure}
\fi
Assuming equal probability $0.5$ for both symbols $+1$ and $-1$ of the BPSK constellation, the transition from the interference-free state $[0, 0, +1]$ into one of the states $[0, 0, +1]$ or $[0, 0, -1]$ occurs with probability $q=\frac{1}{2}(1-p_a)$, since $(1-p_a)$ is the probability that no interferer arrives. The transition from state $[2, 0, -1]$ into one of the states $[2, 1, -1]$ or $[2, 1, +1]$ occurs with probability $p=\frac{1}{2}p_a$. All other transition probabilities are obtained similarly. Note that maximum one interferer can arrive per symbol interval according to the underlying Poisson process assumption.
\subsection{A-posteriori probability calculation}
\label{app_alg}
For the a-posteriori probability calculation, we utilize a MAP detector based on the well-known BCJR algorithm \cite{1055186} which provides a-posteriori probabilities for Markov processes and minimizes the probability of a detection error. The BCJR algorithm is based on a forward recursion and a backward recursion through a trellis diagram obtained from the state transition diagram of the Markov process. The algorithm utilizes the following matrices.
\begin{itemize}
\item Transition matrix $\textbf{P}(m)$, which contains the a-priori transition probabilities $\mathrm{Pr}(s[m]\ |\ s[m-1])$ of the Markov process which are assumed to be identical for different time steps $m$ corresponding to data transmission;
\item likelihood matrix $\textbf{R}$, which contains the likelihoods $\mathrm{pdf}(y_i[m]\: |\: s[m]),\:\forall m$ characterizing the transmission channel,
\end{itemize}
where $s[m]\in\left(\mathcal{S}_+\cup\mathcal{S}_-\right)$ denotes the state of the Markov process at time step $m$.

The transition matrix $\textbf{P}(m)$ can be directly determined based on the underlying Markov product-chain. As an example, the transition matrix for the Markov product-chain given in Fig. \ref{Markov} is
\begin{equation}
\textbf{P}=
\begin{bmatrix}
q & q & 0 & 0 & q & q & 0 & 0\\
0 & 0 & q & q & 0 & 0 & q & q\\
p & p & 0 & 0 & p & p & 0 & 0\\
0 & 0 & p & p & 0 & 0 & p & p\\
q & q & 0 & 0 & q & q & 0 & 0\\
0 & 0 & q & q & 0 & 0 & q & q\\
p & p & 0 & 0 & p & p & 0 & 0\\
0 & 0 & p & p & 0 & 0 & p & p
\end{bmatrix},
\end{equation}
where the columns (and rows) are associated with the states [0, 0, +1], [1, 0, +1], [2, 0, +1], [2, 1, +1], [0, 0, -1], [1, 0, -1], [2, 0, -1], [2, 1, -1] in the given order.

Since the interference is assumed to follow a normal distribution, the likelihoods are given by
\begin{eqnarray}
\hspace*{-3mm}\mathrm{pdf}(y_i[m]\: |\: s[m]\in \mathcal{S}_+)\hspace*{-2.5mm}&=&\hspace*{-2.5mm}\frac{\exp\left(-\frac{(y_i[m]-1)^2}{2\sigma_{\mathrm{tot}}^2(s[m])}\right)}{\sqrt{2\pi\sigma_{\mathrm{tot}}^2(s[m])}},\\
\hspace*{-3mm}\mathrm{pdf}(y_i[m]\: |\: s[m]\in \mathcal{S}_-)\hspace*{-2.5mm}&=&\hspace*{-2.5mm}\frac{\exp\left(-\frac{(y_i[m]+1)^2}{2\sigma_{\mathrm{tot}}^2(s[m])}\right)}{\sqrt{2\pi\sigma_{\mathrm{tot}}^2(s[m])}},
\end{eqnarray}
where $\sigma_{\mathrm{tot}}^2(s[m])$ represents the total variance of the disturbance, which includes the noise variance and the variance of active interferers according to the new state. As an example, a transition emerging from $s[m-1]=\left[4, 3, 1, 0, +1\right]$ is considered. If no additional interferer arrives, the new state will be $s[m]=\left[3, 2, 0, 0, +1\right]$ or $s[m]=\left[3, 2, 0, 0, -1\right]$. In both cases, $\sigma_{\mathrm{tot}}^2(s[m])=\sigma_N^2+2\cdot \sigma_I^2$, since there are only two active interferers. If an additional interferer arrives, the new state will be $s[m]=\left[4, 3, 2, 0, +1\right]$ or $s[m]=\left[4, 3, 2, 0, -1\right]$ which leads to $\sigma_{\mathrm{tot}}^2(s[m])=\sigma_N^2+3\cdot \sigma_I^2$. In fact, the number of active interferers corresponds to the number of non-zero interferer substates in $s[m]$. Thus, all pdfs $\mathrm{pdf}(y_i[m]\: |\: s[m]),\:\forall m$ are determined and stored in matrix $\textbf{R}$.

The BCJR algorithm obtains joint pdfs $\lambda(s[m])=\mathrm{Pr}\left(y_i[\cdot],\:s[m]\right)$ for each state of the Markov chain, where $y_i[\cdot]$ stands for the whole received sub-packet. Then, the a-posteriori probabilities $\mathrm{Pr}\left(x_i[m]=+1\:|\:y_i[\cdot]\right)$ and $\mathrm{Pr}\left(x_i[m]=-1\:|\:y_i[\cdot]\right)$ are determined via summation of all joint pdfs $\lambda(s[m])$, $s[m] \in \mathcal{S}_+$ and $\lambda(s[m])$, $s[m] \in \mathcal{S}_-$, respectively, i.e.,
\ifCLASSOPTIONdraftcls
\begin{eqnarray}
\mathrm{Pr}\left(x_i[m]=+1\:|\:y_i[\cdot]\right)\hspace*{-2mm}&=&\hspace*{-2mm}\frac{1}{\mathrm{pdf}\left(y_i[\cdot]\right)}\sum_{s[m] \in \mathcal{S}_+}\lambda(s[m]),\\
\mathrm{Pr}\left(x_i[m]=-1\:|\:y_i[\cdot]\right)\hspace*{-2mm}&=&\hspace*{-2mm}\frac{1}{\mathrm{pdf}\left(y_i[\cdot]\right)}\sum_{s[m] \in \mathcal{S}_-}\lambda(s[m]).
\end{eqnarray}
\else
\begin{eqnarray}
\hspace*{-3mm}\mathrm{Pr}\left(x_i[m]=+1\:|\:y_i[\cdot]\right)\hspace*{-2.5mm}&=&\hspace*{-2.5mm}\frac{1}{\mathrm{Pr}\left(y_i[\cdot]\right)}\sum_{\mathcal{S}_+}\lambda(s_+[m]),\\
\hspace*{-3mm}\mathrm{Pr}\left(x_i[m]=-1\:|\:y_i[\cdot]\right)\hspace*{-2.5mm}&=&\hspace*{-2.5mm}\frac{1}{\mathrm{Pr}\left(y_i[\cdot]\right)}\sum_{\mathcal{S}_-}\lambda(s_-[m]).
\end{eqnarray}
\fi
Finally, the LLR values are calculated as 
\begin{equation}
\mathrm{LLR}_i[m]=\log\left(\frac{\mathrm{Pr}(x_i[m]=+1\:|\:y_i[\cdot])}{\mathrm{Pr}(x_i[m]=-1\:|\:y_i[\cdot])}\right).
\end{equation}
\subsection{Training sequence and silent periods}
As already mentioned before, the knowledge of some of the transmitted symbols (training symbols) may not only be used for synchronization purposes but can also help in deducing the current state of the Markov process. Correspondingly, the probability of symbol error can be further reduced. In order to account for the training sequence with $L_{\mathrm{tr}}$ symbols inserted in the middle of each sub-packet, we modify the entries of matrix $\textbf{P}(m)$ for time indices $m$ corresponding to the training sequence such that only the transitions from $s[m-1]\in\mathcal{S}_{r[m-1]}$ to $s[m]\in\mathcal{S}_{r[m]}$ are possible and all other entries of the matrix are set to zero. Here, $\mathcal{S}_{r[m]}$ corresponds to either $\mathcal{S}_{+}$ or $\mathcal{S}_{-}$ depending on the current symbol $r[m]$.

Furthermore, due to a typically discontinuous transmission in LPWA networks, we can assume that no data is transmitted shortly before and after each sub-packet. This situation can be exploited in order to further improve the detection performance via a long-term observation of the interference without additional uncertainty resulting from the superposition of the unknown desired signal. For this, we assume that $2L_{\mathrm{add}}$ additional symbols $y_i[m]$ in the intervals $-L_{\mathrm{add}}\leq m<0$ and $L_{\mathrm{tot}}\leq m<L_{\mathrm{tot}}+L_{\mathrm{add}}$ have been collected at the receiver. Since no data is transmitted during these time intervals, we create an additional set of states $\mathcal{S}_0$ which correspond to a transmit symbol $x_i[m]=0$. Hence, during the first and the last $L_{\mathrm{add}}$ considered symbol intervals, the transition matrix $\textbf{P}(m)$ contains the conditional probabilities\footnote{We assume that there are more than $L_{\mathrm{add}}$ symbols at each end of the observed sequence, for which $s[m]\in\mathcal{S}_0$ holds. However, only $L_{\mathrm{add}}$ symbols are taken into account in the calculation. Correspondingly, the transition from $s[-L_{\mathrm{add}}-1]$ to $s[-L_{\mathrm{add}}]$ and from $s[L_{\mathrm{tot}}+L_{\mathrm{add}}-1]$ to 
$s[L_{\mathrm{tot}}+L_{\mathrm{add}}]$
can still be described in this way.} $\mathrm{Pr}(s[m] \: |\: s[m-1])$,
$s[m-1]\in \mathcal{S}_0$, $s[m]\in \mathcal{S}_0$,
and the likelihood matrix $\textbf{R}$ comprises the likelihoods $\mathrm{pdf}(y_i[m]\: |\: s[m])=\frac{1}{\sqrt{2\pi\sigma_{\mathrm{tot}}^2(s[m])}}\exp\left(-\frac{y_i^2[m]}{2\sigma_{\mathrm{tot}}^2(s[m])}\right)$, $s[m]\in \mathcal{S}_0$.

With increasing number of considered symbols $2L_{\mathrm{add}}$, the detection performance can be substantially improved, cf.\ also Section \ref{sim_section}, since more information about the interference and the 
initial and terminal state of the Markov chain, respectively, is obtained, i.e., the degradation due to a missing termination of the trellis diagram at both ends is reduced.

Since the proposed scheme corresponds to the optimal MAP symbol detector, its performance constitutes an upper bound for the performance of any realizable detector. 
\subsection{Multiple classes of interferers}
\label{sec:mult}
For the algorithm according to Section \ref{app_alg}, a constant interference length and variance is assumed. In practice, the interference length can vary, especially if multiple communication systems share the same resources. Also, the interference variance at the receiver is influenced by fading effects and the positions of the nodes within the deployment field. Furthermore, the imperfections of time and frequency synchronization between the LPWA transmitter and receiver have not been considered so far. The synchronization errors may affect the interference variance and length as well. All mentioned phenomena can be accounted for, if multiple classes of interferers are included in the system model and regarded in detection. For example, a certain interferer class may be used in order to describe the interference in case of perfect synchronization, while another class may characterize the interference in the presence of synchronization errors.

Multiple classes of interferers can be modeled via a Markov product-chain resulting from the combination of multiple Markov 
(sub-)chains. Here, each interferer class is characterized by the interference length, variance, and load. Obviously, for each state of one sub-chain, any state of another sub-chain can occur. Hence, the state of the product-chain is defined as a concatenation of the states of the individual sub-chains. The transition probabilities between the states of the product-chain are calculated as a product of the corresponding transition probabilities of the respective states of the sub-chains. We refer to this method as a full-state solution.

The main drawback of this strategy is that the number of states of the resulting product-chain is given by the product of the numbers of states of the individual sub-chains (without taking into account the extension of the state definition with BPSK symbols), i.e., $2^{\sum_{i=1}^K L_{I, i}}$ states are considered, where $L_{I, i}$ is the interferer length of class $i$ 
and $K$ denotes the number of classes. Correspondingly, only few classes (typically 3 or 4) can be included if the computational complexity has to be kept moderate. On the other hand, the number of classes may be high in practical applications. Interestingly, in case of low interference load, it is not highly likely that multiple interferers of the same class would be active simultaneously in order for their transmissions to overlap. If we exclude such overlaps within the same class from being considered, 
the number of states per class (and per sub-chain) reduces from $2^{L_{I, i}}$ to $L_{I, i}+1$. We refer to this method as a reduced-state solution for which the number of states of the product-chain is given by $\prod_{i=1}^K \left(L_{I, i}+1\right)$. Through this, even interferer lengths in the order of the size of the sub-packet can be incorporated which is typically not possible using the full-state method considering all theoretically possible interferer combinations per class due to a too high state number. 

In the following, an example of a reduced-state product-chain with $K=2$ interferer classes of lengths $L_{I, 1}=2$ and $L_{I, 2}=1$ is discussed. The
corresponding state transition matrix for the data transmission phase is given by\footnote{We omit the graphical representation of this Markov chain due to the high number of transitions.}
\ifCLASSOPTIONdraftcls
\begin{equation}
\textbf{P}(m)=
\begin{bmatrix}
\textbf{P}_{\mathrm{small}} & \textbf{P}_{\mathrm{small}}\\
\textbf{P}_{\mathrm{small}} & \textbf{P}_{\mathrm{small}}
\end{bmatrix} \: \text{ with } \: 
\textbf{P}_{\mathrm{small}}=
\begin{bmatrix}
q_1 & q_1 & q_1 & q_1 & 0 & 0\\
q_2 & q_2 & q_2 & q_2 & 0 & 0\\
0 & 0 & 0 & 0 & q_3 & q_3\\
0 & 0 & 0 & 0 & q_4 & q_4\\
q_5 & q_5 & q_5 & q_5 & 0 & 0\\
q_6 & q_6 & q_6 & q_6 & 0 & 0\\
\end{bmatrix},
\end{equation}
\else
\begin{eqnarray}
\textbf{P}(m)\hspace*{-2mm}&=&\hspace*{-2mm}
\begin{bmatrix}
\textbf{P}_{\mathrm{small}} & \textbf{P}_{\mathrm{small}}\\
\textbf{P}_{\mathrm{small}} & \textbf{P}_{\mathrm{small}}
\end{bmatrix}\\ \text{ with } \: 
\textbf{P}_{\mathrm{small}}\hspace*{-2mm}&=&\hspace*{-2mm}
\begin{bmatrix}
q_1 & q_1 & q_1 & q_1 & 0 & 0\\
q_2 & q_2 & q_2 & q_2 & 0 & 0\\
0 & 0 & 0 & 0 & q_3 & q_3\\
0 & 0 & 0 & 0 & q_4 & q_4\\
q_5 & q_5 & q_5 & q_5 & 0 & 0\\
q_6 & q_6 & q_6 & q_6 & 0 & 0\\
\end{bmatrix},\notag
\end{eqnarray}
\fi
where $q_1=\frac{1}{2}(1-p_{a, 1})(1-p_{a, 2})$, $q_2=\frac{1}{2}(1-p_{a, 1})p_{a, 2}$, $q_3=\frac{1}{2}(1-p_{a, 2})$, $q_4=\frac{1}{2}p_{a, 2}$, $q_5=\frac{1}{2}p_{a, 1}(1-p_{a, 2})$, $q_6=\frac{1}{2}p_{a, 1}p_{a, 2}$. Here, the columns (and rows) of matrix $\textbf{P}(m)$ correspond to the states [0,0,+1], [0,1,+1], [1,0,+1], [1,1,+1], [2,0,+1], [2,1,+1], [0,0,-1], [0,1,-1], [1,0,-1], [1,1,-1], [2,0,-1], [2,1,-1], assuming that no further interferer packet of length $L_{I, 1}=2$ can arrive during the occurrence of an active interferer from the first class. This assumption is well justified in case of a relatively low interference load for this class of interferers, i.e., $G_{1}\leq 0.2$. The resulting overall product-chain has $2\left(2+1\right)\left(1+1\right)=12$ states. On the contrary, $2\cdot 2^{2}\cdot 2^{1}=16$ states would have to be utilized if overlaps among the interferers of the first class were accounted for. With more classes and longer interference packets, the number of states in a full-state solution may become even by orders of magnitude larger than that in the advocated reduced-state solution.
\subsection{Further complexity reduction}
\label{sec:red}
Although the reduced-complexity method described in Section \ref{sec:mult} provides a reliable symbol detection in case of a relatively low interference load, its computational complexity might still grow prohibitively large for an increasing number of interference classes, especially if the interference length is not small. Furthermore, it might be very difficult to obtain accurate a-priori information about the characteristics of the interference from coexisting communication systems. In fact, the received signal variance can be hardly estimated accurately even by the receiver of the interfering communication system. Correspondingly, the statistical modeling of the overall received interference might be inaccurate and result in heavy losses in terms of packet error rate. In order to cope with both problems, i.e., a high detector complexity and lack of a priori
knowledge of the full interference characteristics, we propose a detector with scalable complexity which is based on a long-term observation of the interference.

Assuming a discontinuous transmission of data packets by the transmitter, the received signals before and between the actual packet transmissions can be recorded and used in order to estimate the interference variance in the absence of the useful signal. The sequence of squared signal magnitude samples is considered and its most representative values are determined. For this, the squared magnitude samples are sorted in ascending order and subdivided into partitions using one of the known partitioning or clustering methods, e.g. the expectation-maximization (EM) algorithm \cite{dempster1977maximum}. In this work, we employ the Lloyd's algorithm which is based on centroid Voronoi diagrams \cite{1056489}. More specifically, we apply the algorithm to the logarithmically scaled squared magnitudes, which performs better than the partitioning of the original squared magnitudes as we observed in our simulations. As a result, we obtain $P$ partitions and their centroids, which are used to define states of a Markov chain. If the squared signal magnitude is within the range of a particular partition, we assume that the interference can be characterized by the centroid corresponding to that partition. Hence, the considered sequence of samples is quantized based on the determined partitions. Using the sequence of quantized samples, we calculate the transition probabilities between different states by counting the number of transitions between the respective states and normalizing the result. Hence, a Markov chain is determined, which describes the transitions between the states of the interference variance, see Fig. \ref{Markov_scalable}.
\ifCLASSOPTIONdraftcls
\begin{figure}
\centering
\psfrag{s0+}{\begin{normalsize}$[\sigma_{I, 0}^2, +1]$\end{normalsize}}
\psfrag{s1+}{\begin{normalsize}$[\sigma_{I, 1}^2, +1]$\end{normalsize}}
\psfrag{s2+}{\begin{normalsize}$[\sigma_{I, 2}^2, +1]$\end{normalsize}}
\psfrag{s0-}{\begin{normalsize}$[\sigma_{I, 0}^2, -1]$\end{normalsize}}
\psfrag{s1-}{\begin{normalsize}$[\sigma_{I, 1}^2, -1]$\end{normalsize}}
\psfrag{s2-}{\begin{normalsize}$[\sigma_{I, 2}^2, -1]$\end{normalsize}}
\psfrag{p00}{\begin{small}$p_{0\rightarrow 0}$\end{small}}
\psfrag{p01}{\begin{small}$p_{0\rightarrow 1}$\end{small}}
\psfrag{so}{\begin{small}$p_{1\rightarrow 1}$\end{small}}
\psfrag{p10}{\begin{small}$p_{1\rightarrow 0}$\end{small}}
\psfrag{p02}{\begin{small}$p_{0\rightarrow 2}$\end{small}}
\psfrag{p22}{\begin{small}$p_{2\rightarrow 2}$\end{small}}
\psfrag{p20}{\begin{small}$p_{2\rightarrow 0}$\end{small}}
\psfrag{p12}{\begin{small}$p_{1\rightarrow 2}$\end{small}}
\psfrag{p21}{\begin{small}$p_{2\rightarrow 1}$\end{small}}
\includegraphics[width=0.77\textwidth]{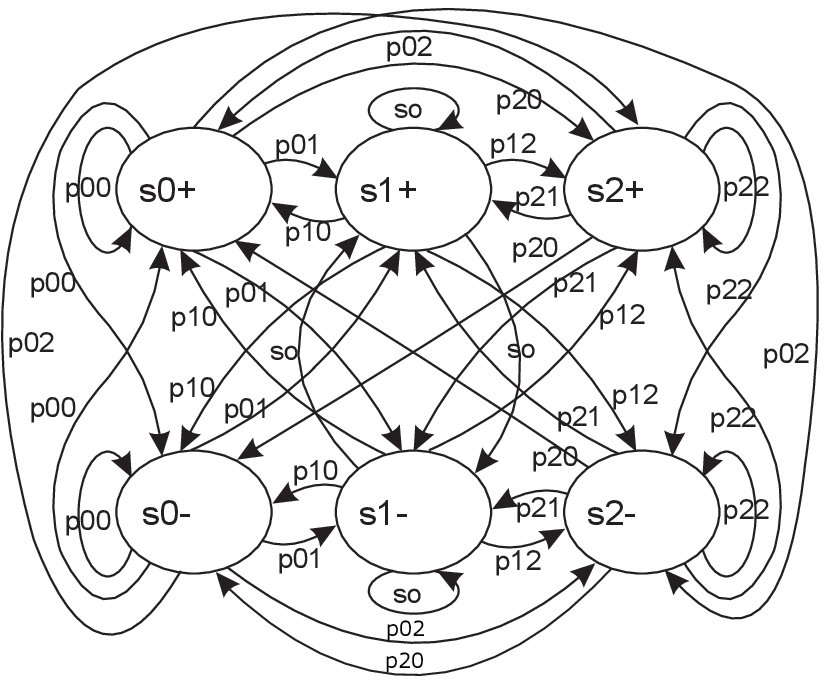}
\caption{A Markov chain for the scalable approach with three states per symbol. Here, the three partitions have centroid variances $\sigma_{I, 0}^2$, $\sigma_{I, 1}^2$, and $\sigma_{I, 2}^2$, respectively. The transition probabilities $p_{s[m]\rightarrow s[m+1]},\forall s[m], m$ are obtained from the measurements.}
\label{Markov_scalable}
\vspace*{-2mm}
\end{figure}
\else
\begin{figure*}[!t]
\centering
\psfrag{s0+}{\begin{normalsize}\hspace*{2mm}$[0, +1]$\end{normalsize}}
\psfrag{s1+}{\begin{normalsize}$[\sigma_{I, 1}^2, +1]$\end{normalsize}}
\psfrag{s2+}{\begin{normalsize}$[\sigma_{I, 2}^2, +1]$\end{normalsize}}
\psfrag{s0-}{\begin{normalsize}\hspace*{2mm}$[0, -1]$\end{normalsize}}
\psfrag{s1-}{\begin{normalsize}$[\sigma_{I, 1}^2, -1]$\end{normalsize}}
\psfrag{s2-}{\begin{normalsize}$[\sigma_{I, 2}^2, -1]$\end{normalsize}}
\psfrag{p00}{\begin{small}$p_{0\rightarrow 0}$\end{small}}
\psfrag{p01}{\begin{small}$p_{0\rightarrow 1}$\end{small}}
\psfrag{so}{\begin{small}$p_{1\rightarrow 1}$\end{small}}
\psfrag{p10}{\begin{small}$p_{1\rightarrow 0}$\end{small}}
\psfrag{p02}{\begin{small}$p_{0\rightarrow 2}$\end{small}}
\psfrag{p22}{\begin{small}$p_{2\rightarrow 2}$\end{small}}
\psfrag{p20}{\begin{small}$p_{2\rightarrow 0}$\end{small}}
\psfrag{p12}{\begin{small}$p_{1\rightarrow 2}$\end{small}}
\psfrag{p21}{\begin{small}$p_{2\rightarrow 1}$\end{small}}
\includegraphics[width=0.65\textwidth]{kisse6.eps}
\caption{A Markov chain for the scalable approach with three states per symbol. Here, the three partitions have centroid variances $\sigma_{I, 0}^2$, $\sigma_{I, 1}^2$, and $\sigma_{I, 2}^2$, respectively. The transition probabilities $p_{s[m]\rightarrow s[m+1]},\forall s[m], m$ are obtained from the measurements.}
\label{Markov_scalable}
\vspace*{-2mm}
\end{figure*}
\fi
An alternative representation of this Markov chain via state transition matrix is given below:
\begin{equation}
\textbf{P}=
\begin{bmatrix}
\textbf{P}_{\mathrm{small}} & \textbf{P}_{\mathrm{small}}\\
\textbf{P}_{\mathrm{small}} & \textbf{P}_{\mathrm{small}}
\end{bmatrix},\:
\textbf{P}_{\mathrm{small}}=
\begin{bmatrix}
p_{0\rightarrow 0} & p_{1\rightarrow 0} & p_{2\rightarrow 0}\\
p_{0\rightarrow 1} & p_{1\rightarrow 1} & p_{2\rightarrow 1}\\
p_{0\rightarrow 2} & p_{1\rightarrow 2} & p_{2\rightarrow 2}
\end{bmatrix},
\end{equation}
where the columns (and rows) of $\textbf{P}$ correspond to the following states: [$\sigma_{I, 0}^2$, +1],  [$\sigma_{I, 1}^2$, +1],  [$\sigma_{I, 2}^2$, +1],  [$\sigma_{I, 0}^2$, -1],  [$\sigma_{I, 1}^2$, -1],  [$\sigma_{I, 2}^2$, -1]. The transition probabilities $p_{s[m]\rightarrow s[m+1]},\forall s[m], m$ are obtained from the measurements. Using the transition probabilities of the Markov chain, the detection is done using the BCJR algorithm as discussed before.

A related method has been proposed in \cite{7564982} for symbol detection in the presence of impulsive noise for powerline communications. However, a fixed noise model with a given and constant number of states and known noise variances corresponding to the states and transition probabilities between states has been assumed in \cite{7564982}. Thus, it was sufficient to consider 19 states for the full-state modeling of the impulsive noise. Then, using a BCJR algorithm, the a-posteriori symbol probabilities have been calculated. In contrast, our method does not require any prior knowledge of the interference variances, number of states for the full-state modeling or similar information, since the proposed scalable approach relies solely on the observation of the interference signal. Furthermore, the complexity of the proposed method is very low, since we obtain only $2P$ states of the Markov chain.

The major benefit of this method is a much lower computational complexity of the detector, since the length of the interference packets is not relevant for this approach. Instead, the complexity in terms of the number of states is given by the number of partitions, which is a parameter that can be freely chosen. Correspondingly, we refer to this scheme as "scalable detector" in Section \ref{sec_4}. Another benefit of this scheme is that it is fully blind, i.e., no a-priori knowledge of the interference statistics is required. However, the method is somewhat less accurate compared to the full-state (optimal) MAP detection described in Section \ref{sec_3_1}, where the (available) interference statistics are fully exploited. The accuracy of the scheme can be further improved by refining the transition probabilities using the well-known Baum-Welch algorithm, cf.  \cite{dempster1977maximum}.
\section{Numerical Results}
\label{sec_4}
\label{sim_section}
\subsection{Simulation parameters}
For the simulation results discussed in the following, we have chosen a convolutional code with 6 memory elements and a code rate $R_c=\frac{1}{3}$, where the number of encoder input bits and the number of encoded bits is $k=168$ and $n=504$, respectively. For trellis termination, 6 tail bits are input to the encoder after the information bits. Furthermore, the length of each sub-packet is $L_S=28$ and correspondingly the number of sub-packets per telegram is $n/L_S=18$. A training sequence of length $L_{\mathrm{tr}}=8$ is employed, given by $[-1, -1, -1, +1, -1, +1, +1, +1]$. The symbol duration is selected to $T=0.5$ ms. For the transmit filter $g(t)$, we adopt a cosine half wave of duration $T$ corresponding to minimum-shift keying (MSK) modulation. For each scenario, the packet error rate (PER) is obtained from a simulation of 20000 packet transmissions. A packet error occurs in case of at least one bit error after FEC decoding of the whole received data packet. We show the simulated PER for different signal-to-noise ratios $E_s/N_0=1/\sigma_N^2$ and interference lengths $L_I$, where $E_s$ stands for the energy per transmitted symbol and $N_0$ is the noise power spectral density. It should be noted that the $E_s/N_0$ range adopted in our paper is a practical one since its definition is  related  to  the  ultra-narrow  transmission  band  of  a  single  node  in  telegram  splitting  (which applies frequency hopping in addition) and not to the overall system bandwidth (for which the $E_s/N_0$ values would be much lower).

In order to guarantee the best possible performance of the proposed schemes even with large interference variance, our investigations have shown that the exploitation of $L_{\mathrm{add}}=10$ additional symbols before and after each sub-packet
is sufficient. Hence, this choice is adopted for all results.

For the interference variance, we assume $\sigma_I^2=2$ in scenarios with a single interference class. In scenarios with multiple interference classes, we consider different interference variances, which may represent the signal transmissions from co-located nodes in similar distances from the receiver. Such co-located nodes can be viewed as node clusters.

In this work, we focus on moderate interference loads $G\leq 0.6$ in order to exploit the benefits of the TS method. Larger loads, e.g. $G\approx 1$, indicate that a new interfering packet arrives in almost every symbol interval. Correspondingly, the interference variance remains almost constant in all consecutive symbol intervals, which is in contradiction to the assumption of bursty interference. Besides, such loads result in a poor performance since a low receive signal quality prevails throughout each code word, and corrupted symbols cannot be corrected anymore via FEC.
\subsection{Baseline schemes}
For comparison, we show also results for genie-aided detection, a technique from \cite{5504595} based on the erasure of corrupted symbols, and a naive approach with assumed constant interference variance, respectively.

For genie-aided detection, the total variance of interference plus noise $\sigma_{\mathrm{tot}, i}^2[m]$ is assumed to be known, such that the LLRs can be expressed as
\ifCLASSOPTIONdraftcls
\begin{equation}
\label{sec4eq1}
\mathrm{LLR}_i[m]=\log\left(\frac{\mathrm{Pr}(x_i[m]=+1\:|\:y_i[\cdot])}{\mathrm{Pr}(x_i[m]=-1\:|\:y_i[\cdot])}\right)=\log\left(\frac{\mathrm{Pr}(y_i[m]\: |\: x_i[m]=+1)}{\mathrm{Pr}(y_i[m]\: |\: x_i[m]=-1)}\right)=\frac{2y_i[m]}{\sigma_{\mathrm{tot}, i}^2[m]}.
\end{equation}
\else
\begin{eqnarray}
\label{sec4eq1}
\hspace*{-5mm}\mathrm{LLR}_i[m]\hspace*{-2.5mm}&=&\hspace*{-2.5mm}\log\left(\frac{\mathrm{Pr}(x_i[m]=+1\:|\:y_i[\cdot])}{\mathrm{Pr}(x_i[m]=-1\:|\:y_i[\cdot])}\right)\notag\\
\hspace*{-5mm}\hspace*{-2.5mm}&=&\hspace*{-2.5mm}\log\left(\frac{\mathrm{Pr}(y_i[m]\: |\: x_i[m]=+1)}{\mathrm{Pr}(y_i[m]\: |\: x_i[m]=-1)}\right)\notag\\
\hspace*{-2.5mm}&=&\hspace*{-2.5mm}\frac{2 \, y_i[m]}{\sigma_{\mathrm{tot}, i}^2[m]}.
\end{eqnarray}
\fi
Unlike other benchmark schemes, the genie-aided detector is not a practical method since it assumes the perfect knowledge of the interference variance in  each symbol interval which is unknown in practice. However, its performance represents a theoretical upper bound for any realizable detector and thus can be used for performance evaluations. Accordingly, analytical expressions for its PER performance are of interest. Corresponding semi-analytical results are provided in the Appendix based on the concept of the {\em effective} signal-to-interference-and-noise ratio (SINR). According to the analysis in the Appendix, for a given realization of interference variances within a codeword, the PER of the genie-aided detector can be approximated as 
\begin{equation}
    {\rm PER} \approx \Psi \left( {\rm SINR}_{\rm eff} \right),
    \label{performance_bound_ga}
\end{equation}
where ${\rm SINR}_{\rm eff}$ is given by (\ref{eq_SINR_eff_1}), (\ref{eq_SINR_eff_2}) and $\Psi (\cdot)$ stands for a nonlinear function.
Here, $\Psi \left( E_s / N_0 \right)$ gives the PER of the adopted convolutional code for the AWGN with a signal-to-noise ratio of $E_s/N_0$. Thus, $\Psi (\cdot)$ can be determined by simulations for the AWGN channel and is independent of the interference conditions.

According to (\ref{performance_bound_ga}), for comparing the performance of the genie-aided detector under different interference variance realizations, a comparison of the corresponding effective SINRs is sufficient. Finally, for an overall performance approximation, 
(\ref{performance_bound_ga}) can be averaged over the interference statistics.

For the erasure method, the presence of interference has been detected using a BCJR algorithm based on a Markov chain with only two states ("good" and "bad"). Here, the average interference variance and the transition probabilities were estimated via long-term observation of the receive signal prior to the data transmission. The symbols pertaining to the "bad" state are excluded ("erased") from the decoding process. This has been accomplished by setting the respective LLR values to zero.

For the naive approach with assumed constant interference variance, the LLRs are calculated according to \eqref{sec4eq1} with a constant variance $\sigma_{\mathrm{tot}, i}^2[m]=\sigma_{\mathrm{tot}}^2,\forall\: i, m$. Since $\sigma_{\mathrm{tot}}^2$ is a constant factor, which is identical for all computed LLRs, the choice of $\sigma_{\mathrm{tot}}^2$ does not affect the decisions of the Viterbi decoder. Hence, we set $\sigma_{\mathrm{tot}}^2=2$, such that
\begin{equation}
\mathrm{LLR}_i[m]=y_i[m]
\end{equation} 
holds. The obvious drawback of this scheme is that no information regarding the interference statistics is taken into account, which could help to distinguish between the reliable and unreliable symbols. In the presence of strong interference, the resulting receive symbols can have a large magnitude, i.e., significantly larger than the magnitude of the useful signal, which indicates that the symbol is unreliable. However, this naive approach would interpret such symbols as more reliable than the received symbols with small disturbance, thus well overlapping with the symbol constellation.

For clarity of notation, we introduce some abbreviations for the considered detectors, see Table \ref{table_abbr}.
\begin{table}[]
\caption{Detectors abbreviations}
    \label{table_abbr}
    \centering
    \begin{tabular}{|c|c|}
    \hline
        \textbf{Detector} & \textbf{Abbreviation} \\
        \hline
        genie-aided detector & genie-aided\\
        \hline
        erasure method & erasure\\
        \hline
        constant variance detector & const. var.\\
        \hline
        full MAP detector & MAP \textit{or} MAP, full\\
        \hline
        reduced-complexity MAP detector & MAP, reduced\\
        \hline
        scalable-complexity MAP detector & MAP, scalable\\
        \hline
    \end{tabular}
\end{table}
These abbreviations will be used in the figures to clearly distinguish between the various detectors.
\subsection{Single interference class}
In Fig.\ \ref{Nguard}, PER vs.\ $ E_s/N_0$ is shown for the full-state MAP detector and the considered baseline schemes for $L_I=6$, $G=0.5$.    
\ifCLASSOPTIONdraftcls
\begin{figure}
\centering
\psfrag{SNR}{\hspace*{-5mm}\begin{normalsize}$E_s/N_0$ [dB]\end{normalsize}}
\psfrag{Packet error rate}{\begin{normalsize}Packet error rate\end{normalsize}}
\psfrag{genie}{\begin{footnotesize}genie-aided\end{footnotesize}}
\psfrag{erasure}{\begin{footnotesize}erasure\end{footnotesize}}
\psfrag{const}{\begin{footnotesize}const. var.\end{footnotesize}}
\psfrag{MAPLadd0}{\begin{footnotesize}MAP\end{footnotesize}}
\psfrag{approx}{\begin{footnotesize}approx.\end{footnotesize}}
\includegraphics[width=0.7\textwidth]{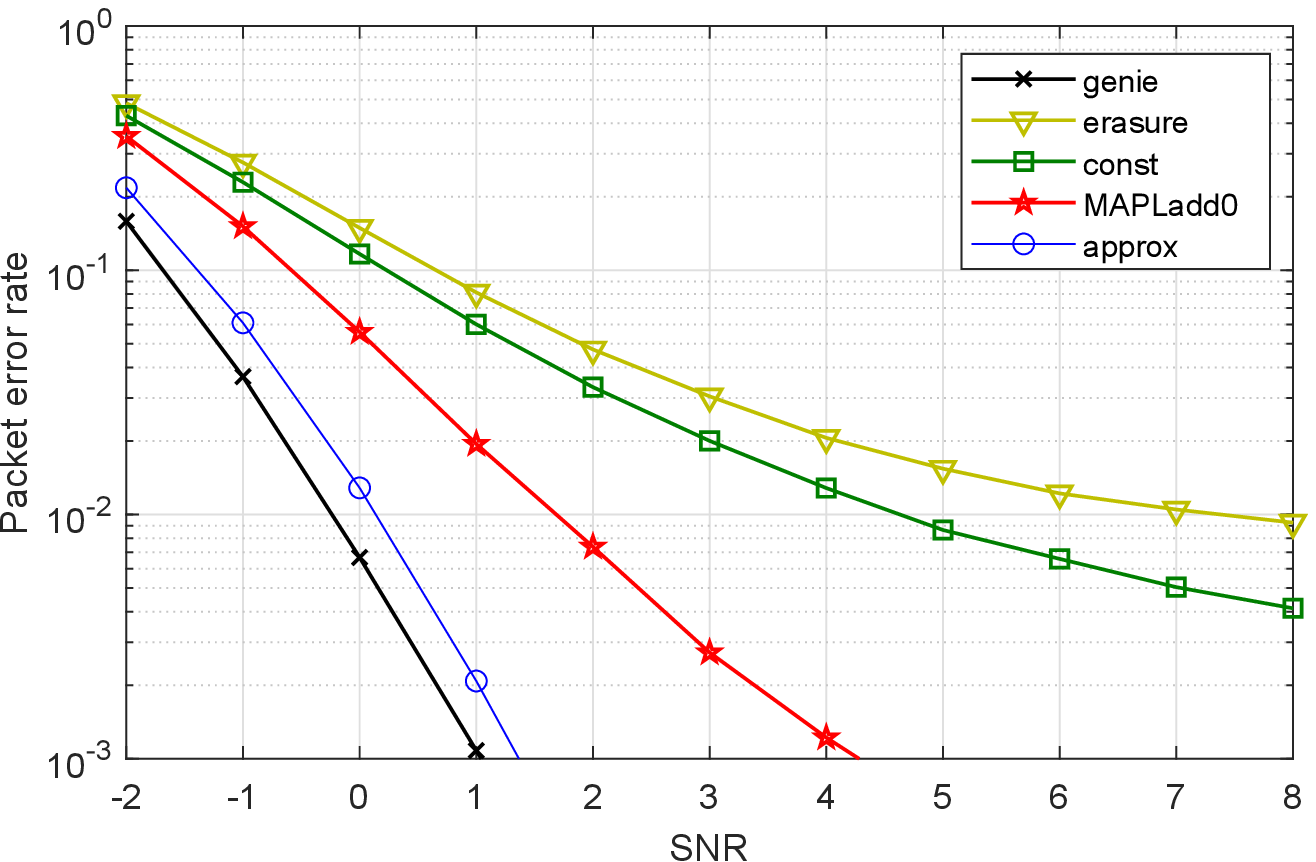}
\caption{Decoding performance of the baseline schemes and the full-state MAP detector for $L_I=6$ and $G=0.5$.}
\label{Nguard}
\vspace*{-2mm}
\end{figure}
\else
\begin{figure}
\psfrag{SNR}{\begin{small}\hspace*{-4mm}$E_s/N_0$ [dB]\end{small}}
\psfrag{Packet error rate}{\begin{small}Packet error rate\end{small}}
\psfrag{genie}{\begin{scriptsize}genie-aided\end{scriptsize}}
\psfrag{erasure}{\begin{scriptsize}erasure\end{scriptsize}}
\psfrag{const}{\begin{scriptsize}const. var.\end{scriptsize}}
\psfrag{MAPLadd0}{\begin{scriptsize}MAP\end{scriptsize}}
\psfrag{approx}{\begin{scriptsize}approx.\end{scriptsize}}
\includegraphics[width=0.49\textwidth]{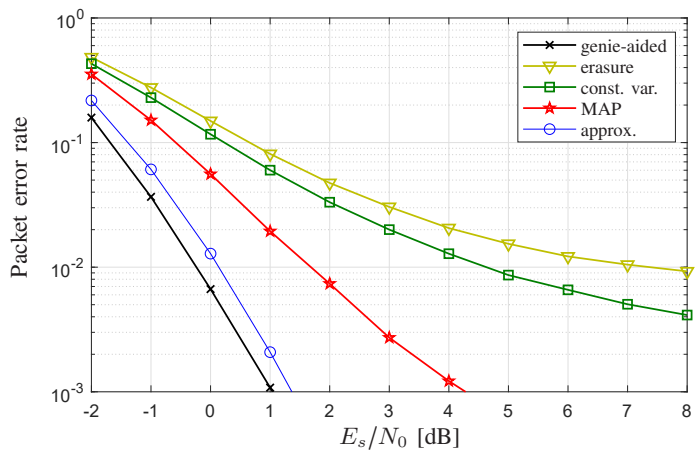}
\caption{Decoding performance of the baseline schemes and the full-state MAP detector for $L_I=6$ and $G=0.5$.}
\label{Nguard}
\vspace*{-2mm}
\end{figure}
\fi
We observe that both the erasure method and the naive approach assuming a constant interference variance perform significantly worse than the proposed full-state MAP detector. Both schemes do not reach $\mathrm{PER}=10^{-3}$ even at very high $E_s/N_0$ values.
Furthermore, we note that there is a performance gap between the full-state MAP detector and the genie-aided detector representing an upper performance bound of approximately 3 dB at $\mathrm{PER}=10^{-3}$. It can be also observed from Fig.\ \ref{Nguard} that the performance approximation according to (\ref{performance_bound_ga}) ('approx.') is quite accurate. Hence, it enables a quick semi-analytical determination of the performance of the genie-aided detector. In the following, we restrict ourselves to simulation results for the genie-aided detector.

For a better insight, we show results for a constant interference load $G=0.5$ and various interferer lengths in Fig. \ref{load}.
\ifCLASSOPTIONdraftcls
\begin{figure}
\centering
\psfrag{SNR}{\hspace*{-5mm}\begin{normalsize}$E_s/N_0$ [dB]\end{normalsize}}
\psfrag{Packet error rate}{\begin{normalsize}Packet error rate\end{normalsize}}
\psfrag{genie2}{\begin{footnotesize}genie-aided, $L_I=3$\end{footnotesize}}
\psfrag{genie6}{\begin{footnotesize}genie-aided, $L_I=6$\end{footnotesize}}
\psfrag{erasure2}{\begin{footnotesize}erasure, $L_I=3$\end{footnotesize}}
\psfrag{erasure6}{\begin{footnotesize}erasure, $L_I=6$\end{footnotesize}}
\psfrag{const2}{\begin{footnotesize}const. var., $L_I=3$\end{footnotesize}}
\psfrag{const6}{\begin{footnotesize}const. var., $L_I=6$\end{footnotesize}}
\psfrag{MAP2}{\begin{footnotesize}MAP, $L_I=3$\end{footnotesize}}
\psfrag{MAP6}{\begin{footnotesize}MAP, $L_I=6$\end{footnotesize}}
\includegraphics[width=0.7\textwidth]{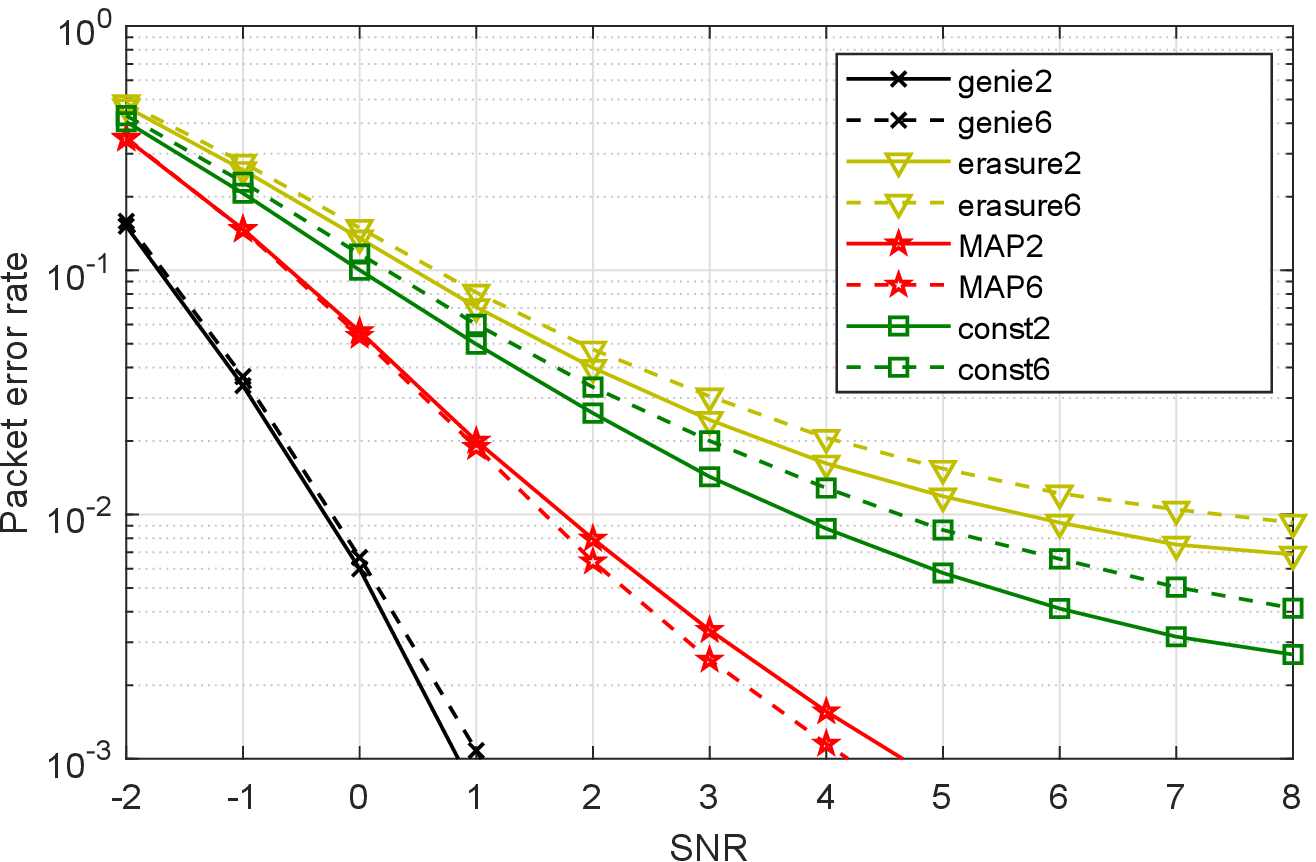}
\caption{Decoding performance of the baseline schemes and the full-state MAP detector for various interferer lengths under constant load $G=0.5$.}
\label{load}
\vspace*{-2mm}
\end{figure}
\else
\begin{figure}
\psfrag{SNR}{\hspace*{-5mm}\begin{small}$E_s/N_0$ [dB]\end{small}}
\psfrag{Packet error rate}{\begin{small}Packet error rate\end{small}}
\psfrag{genie2}{\begin{scriptsize}genie-aided, $L_I=3$\end{scriptsize}}
\psfrag{genie6}{\begin{scriptsize}genie-aided, $L_I=6$\end{scriptsize}}
\psfrag{erasure2}{\begin{scriptsize}erasure, $L_I=3$\end{scriptsize}}
\psfrag{erasure6}{\begin{scriptsize}erasure, $L_I=6$\end{scriptsize}}
\psfrag{const2}{\begin{scriptsize}const. var., $L_I=3$\end{scriptsize}}
\psfrag{const6}{\begin{scriptsize}const. var., $L_I=6$\end{scriptsize}}
\psfrag{MAP2}{\begin{scriptsize}MAP, $L_I=3$\end{scriptsize}}
\psfrag{MAP6}{\begin{scriptsize}MAP, $L_I=6$\end{scriptsize}}
\includegraphics[width=0.49\textwidth]{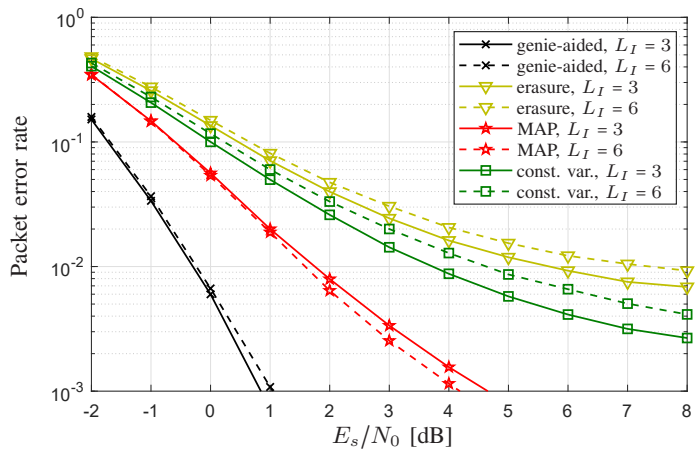}
\caption{Decoding performance of the baseline schemes and the full-state MAP detector for various interferer lengths under constant load $G=0.5$.}
\label{load}
\vspace*{-2mm}
\end{figure}
\fi
Similar to Fig. \ref{Nguard}, the proposed method outperforms the erasure method by at least one order of magnitude for $E_s/N_0\geq3$ dB and $L_I=3$ as well as $L_I=6$. With a further increasing interferer length, the packet length becomes more and more important for an accurate estimation of the a-posteriori probabilities, which is very beneficial for the proposed method. At the same time, more and more packets can potentially overlap, which leads to slightly higher average interference variance, such that the performance of all other methods slightly deteriorates.

Next, we would like to investigate the decoding performance for a constant arrival probability and various lengths of the interferers. For this, we set the ratio $G/L_I$ to a constant value of $0.1$. Hence, $p_a$ would remain constant according to \eqref{eq_2} even if $L_I$ changes ($G$ needs to be adjusted to each scenario). The corresponding results are shown in Fig. \ref{Arrival}.
\ifCLASSOPTIONdraftcls
\begin{figure}
\centering
\psfrag{SNR}{\hspace*{-5mm}\begin{normalsize}$E_s/N_0$ [dB]\end{normalsize}}
\psfrag{Packet error rate}{\begin{normalsize}Packet error rate\end{normalsize}}
\psfrag{genie2}{\begin{footnotesize}genie-aided, $L_I=3$\end{footnotesize}}
\psfrag{genie6}{\begin{footnotesize}genie-aided, $L_I=6$\end{footnotesize}}
\psfrag{erasure2}{\begin{footnotesize}erasure, $L_I=3$\end{footnotesize}}
\psfrag{erasure6}{\begin{footnotesize}erasure, $L_I=6$\end{footnotesize}}
\psfrag{const2}{\begin{footnotesize}const. var., $L_I=3$\end{footnotesize}}
\psfrag{const6}{\begin{footnotesize}const. var., $L_I=6$\end{footnotesize}}
\psfrag{MAP2}{\begin{footnotesize}MAP, $L_I=3$\end{footnotesize}}
\psfrag{MAP6}{\begin{footnotesize}MAP, $L_I=6$\end{footnotesize}}
\includegraphics[width=0.7\textwidth]{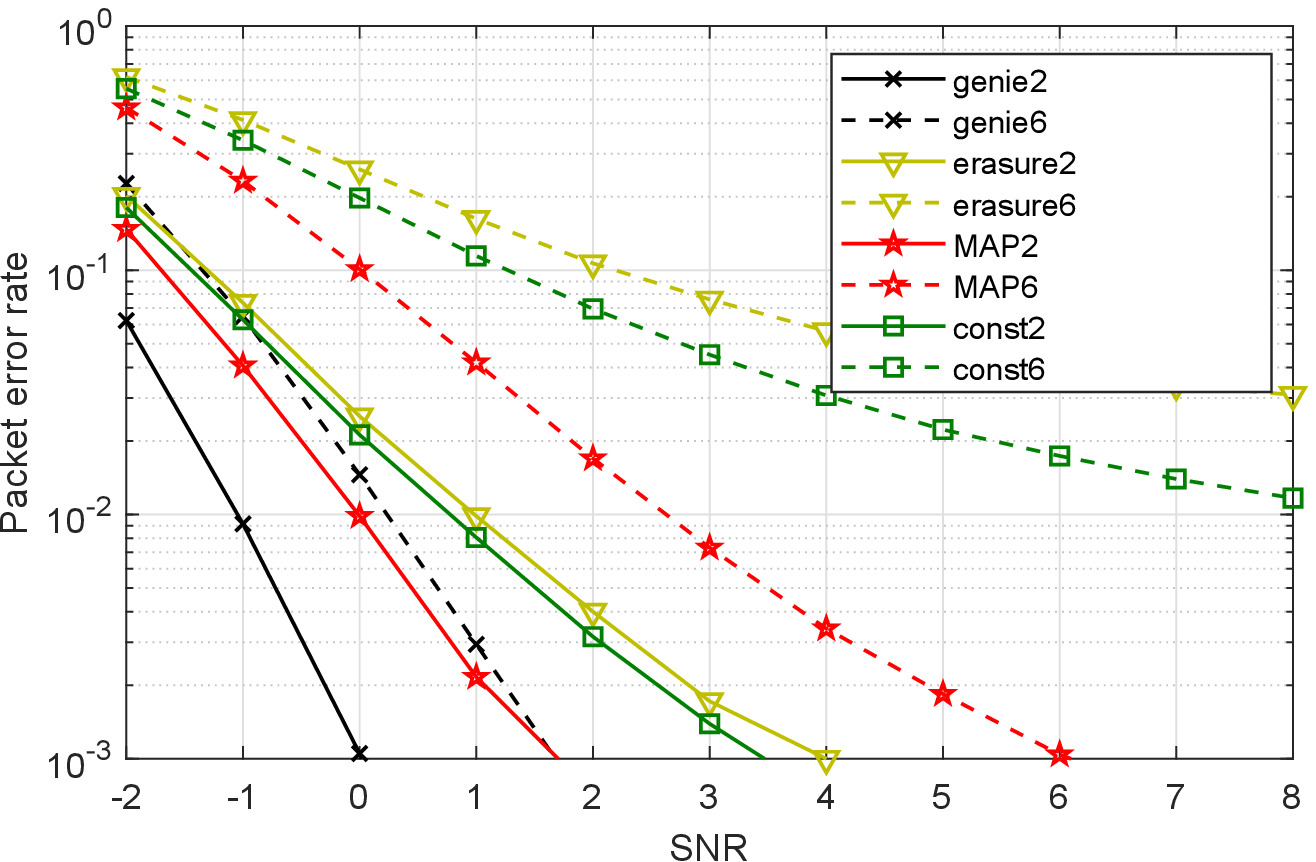}
\caption{Decoding performance of the baseline schemes and the full-state MAP detector and for various interferer lengths under constant arrival rate with $G=0.1\cdot L_I$.}
\label{Arrival}
\vspace*{-2mm}
\end{figure}
\else
\begin{figure}
\psfrag{SNR}{\hspace*{-5mm}\begin{small}$E_s/N_0$ [dB]\end{small}}
\psfrag{Packet error rate}{\begin{small}Packet error rate\end{small}}
\psfrag{genie2}{\begin{scriptsize}genie-aided, $L_I=3$\end{scriptsize}}
\psfrag{genie6}{\begin{scriptsize}genie-aided, $L_I=6$\end{scriptsize}}
\psfrag{erasure2}{\begin{scriptsize}erasure, $L_I=3$\end{scriptsize}}
\psfrag{erasure6}{\begin{scriptsize}erasure, $L_I=6$\end{scriptsize}}
\psfrag{const2}{\begin{scriptsize}const. var., $L_I=3$\end{scriptsize}}
\psfrag{const6}{\begin{scriptsize}const. var., $L_I=6$\end{scriptsize}}
\psfrag{MAP2}{\begin{scriptsize}MAP, $L_I=3$\end{scriptsize}}
\psfrag{MAP6}{\begin{scriptsize}MAP, $L_I=6$\end{scriptsize}}
\includegraphics[width=0.49\textwidth]{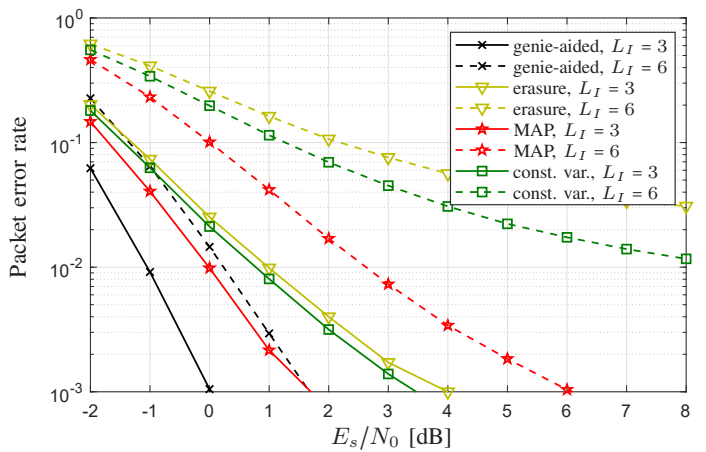}
\caption{Decoding performance of the baseline schemes and the full-state MAP detector and for various interferer lengths under constant arrival rate with $G=0.1\cdot L_I$.}
\label{Arrival}
\vspace*{-2mm}
\end{figure}
\fi
Obviously, with increasing interferer length, PER increases for all  considered schemes. This is related to the fact that the load $G$ increases as well for a constant arrival probability. Correspondingly, more symbols are overlapped by the interfering packets which dramatically affects the decoding performance. Also, we observe that the proposed solution provides a substantially lower PER compared to the erasure method and the constant variance method.
\subsection{Multiple interference classes}
For the analysis of symbol detection with multiple interferer classes, we consider a scenario with two classes which are characterized by packet lengths $L_{I, 1}=2$, $L_{I, 2}=4$ and adaptive load, $G_i=0.2\cdot L_{I, i},\: i\in\{1, 2\}$. The variance of each class is reduced to 1. This scenario describes a communication system with two clusters of interferers located in two different small areas at the same distance from the receiver. The number of clusters is kept low in order to enable also a comparison of the reduced-state MAP solution (according to Section \ref{sec:mult}) with the full-state MAP solution. The results are depicted in Fig.\ \ref{multiclass1}.
\ifCLASSOPTIONdraftcls
\begin{figure}
\centering
\psfrag{SNR}{\hspace*{-5mm}\begin{normalsize}$E_s/N_0$ [dB]\end{normalsize}}
\psfrag{Packet error rate}{\begin{normalsize}Packet error rate\end{normalsize}}
\psfrag{genie}{\begin{footnotesize}genie-aided\end{footnotesize}}
\psfrag{erasure}{\begin{footnotesize}erasure\end{footnotesize}}
\psfrag{const}{\begin{footnotesize}const. var.\end{footnotesize}}
\psfrag{MAP}{\begin{footnotesize}MAP, full\end{footnotesize}}
\psfrag{MAP reduced}{\begin{footnotesize}MAP, reduced\end{footnotesize}}
\includegraphics[width=0.7\textwidth]{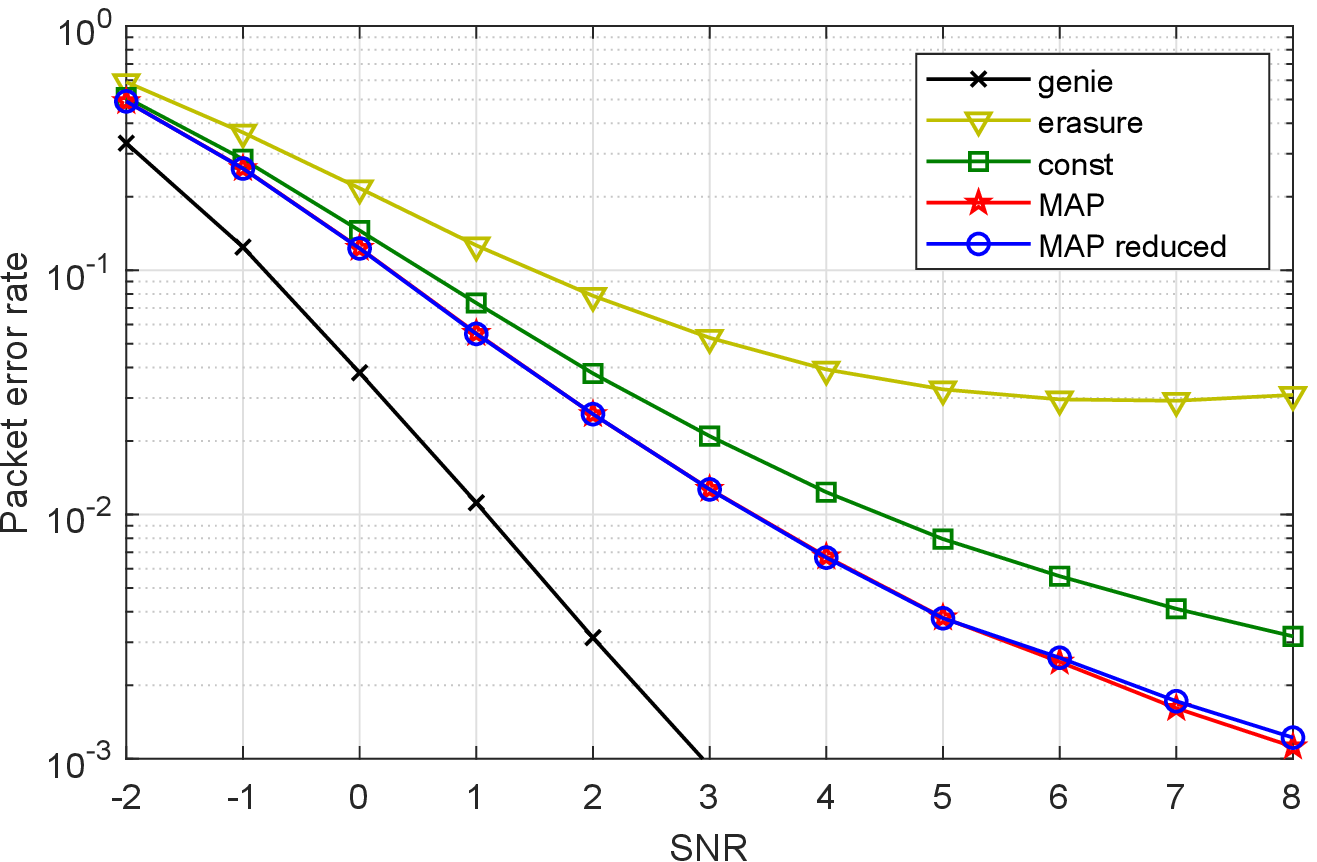}
\caption{Decoding performance of the baseline schemes, the full-state and the reduced-state MAP detectors for a scenario with two interferer classes. $L_{I, 1}=2$, $L_{I, 2}=4$, $G_1=0.4$, $G_2=0.8$, $\sigma_{I, 1}^2=\sigma_{I, 2}^2=1$.}
\label{multiclass1}
\vspace*{-2mm}
\end{figure}
\else
\begin{figure}
\psfrag{SNR}{\hspace*{-5mm}\begin{small}$E_s/N_0$ [dB]\end{small}}
\psfrag{Packet error rate}{\begin{small}Packet error rate\end{small}}
\psfrag{genie}{\begin{scriptsize}genie-aided\end{scriptsize}}
\psfrag{erasure}{\begin{scriptsize}erasure\end{scriptsize}}
\psfrag{const}{\begin{scriptsize}const. var.\end{scriptsize}}
\psfrag{MAP}{\begin{scriptsize}MAP, full\end{scriptsize}}
\psfrag{MAP reduced}{\begin{scriptsize}MAP, reduced\end{scriptsize}}
\includegraphics[width=0.49\textwidth]{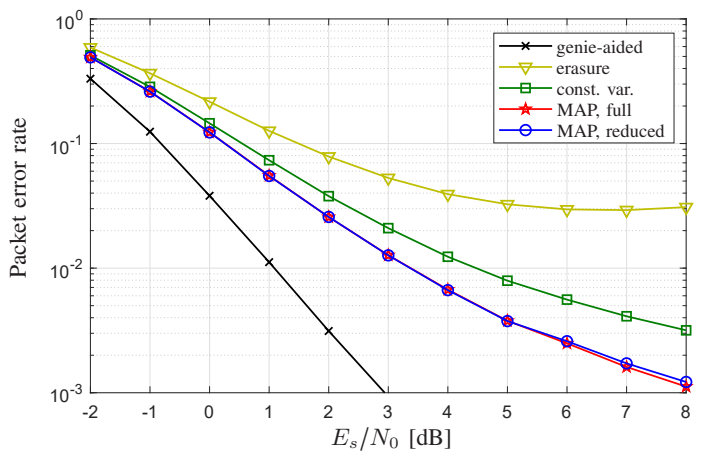}
\caption{Decoding performance of the baseline schemes, the full-state and the reduced-state MAP detectors for a scenario with two interferer classes. $L_{I, 1}=2$, $L_{I, 2}=4$, $G_1=0.4$, $G_2=0.8$, $\sigma_{I, 1}^2=\sigma_{I, 2}^2=1$.}
\label{multiclass1}
\vspace*{-2mm}
\end{figure}
\fi
We observe that the PER is much higher for the erasure method compared to both proposed methods based on MAP detection and to the constant variance approach. Also, the full-state MAP scheme slightly outperforms the reduced-state solution for $E_s/N_0\geq 6$ dB. Nevertheless, the gap between these two schemes is very small due to a moderate interference load for each interferer class, such that the overlaps between multiple transmissions within the same class are still relatively rare. If the load reduces to even lower values, the performance difference between the full-state solution and the reduced-state solution vanishes completely, since overlapping transmissions of the same class are very unlikely. The reduced-state solution should be employed in such scenarios for complexity reasons.

Since the full-state and reduced-state methods rely on the knowledge of the interference variance, it is important to investigate the performance of these schemes in case of imperfect estimation of the interference variance. For this, we assume that the estimation error has a normal distribution with a standard deviation (STD) of 20$\%$, 30$\%$ or 50$\%$ of the true variance. Furthermore, we consider the same scenario as in Fig. \ref{multiclass1}, i.e., two interference classes with $L_{I, 1}=2$, $L_{I, 2}=4$, $G_1=0.4$, and $G_2=0.8$. The results are depicted in Fig. \ref{multiclass2}. For an STD of 20$\%$, both methods show only a small performance degradation compared to the scenario with perfectly known interference variance.
However, we observe a significant performance degradation of both methods with an STD of 30$\%$ at high SNR, which is due to the incorrect calculation of a-posteriori probabilities and LLRs. If we interpret the deviation of the interference variance as an additional distortion, this distortion becomes dominant only in case of low noise variance. In contrast, at low SNR, i.e., for $E_s/N_0\leq 3$ dB, the degradation is negligible. For an STD of 50$\%$, the additional distortion becomes dominant at much lower SNR than for 30$\%$. In fact, both MAP methods fail to reach even PERs of $10^{-2}$ for any SNR in this case. Hence, we conclude that these two methods rely on a sufficiently accurate estimation of the interference variance, preferably with an STD of 30$\%$ or smaller. A sufficiently good estimation of the interference characteristics can be also achieved by the adaptive detector with scalable complexity whose performance is discussed in the following.
\ifCLASSOPTIONdraftcls
\begin{figure}
\centering
\psfrag{SNR}{\hspace*{-5mm}\begin{normalsize}$E_s/N_0$ [dB]\end{normalsize}}
\psfrag{Packet error rate}{\begin{normalsize}Packet error rate\end{normalsize}}
\psfrag{MAP}{\begin{footnotesize}MAP, full, perfect\end{footnotesize}}
\psfrag{Red}{\begin{footnotesize}MAP, reduced, perfect\end{footnotesize}}
\psfrag{MP20}{\begin{footnotesize}MAP, full, 20$\%$ error\end{footnotesize}}
\psfrag{Red20}{\begin{footnotesize}MAP, reduced, 20$\%$ error\end{footnotesize}}
\psfrag{MP30}{\begin{footnotesize}MAP, full, 30$\%$ error\end{footnotesize}}
\psfrag{Red30}{\begin{footnotesize}MAP, reduced, 30$\%$ error\end{footnotesize}}
\psfrag{MP50}{\begin{footnotesize}MAP, full, 50$\%$ error\end{footnotesize}}
\psfrag{Red50}{\begin{footnotesize}MAP, reduced, 50$\%$ error\end{footnotesize}}
\includegraphics[width=0.7\textwidth]{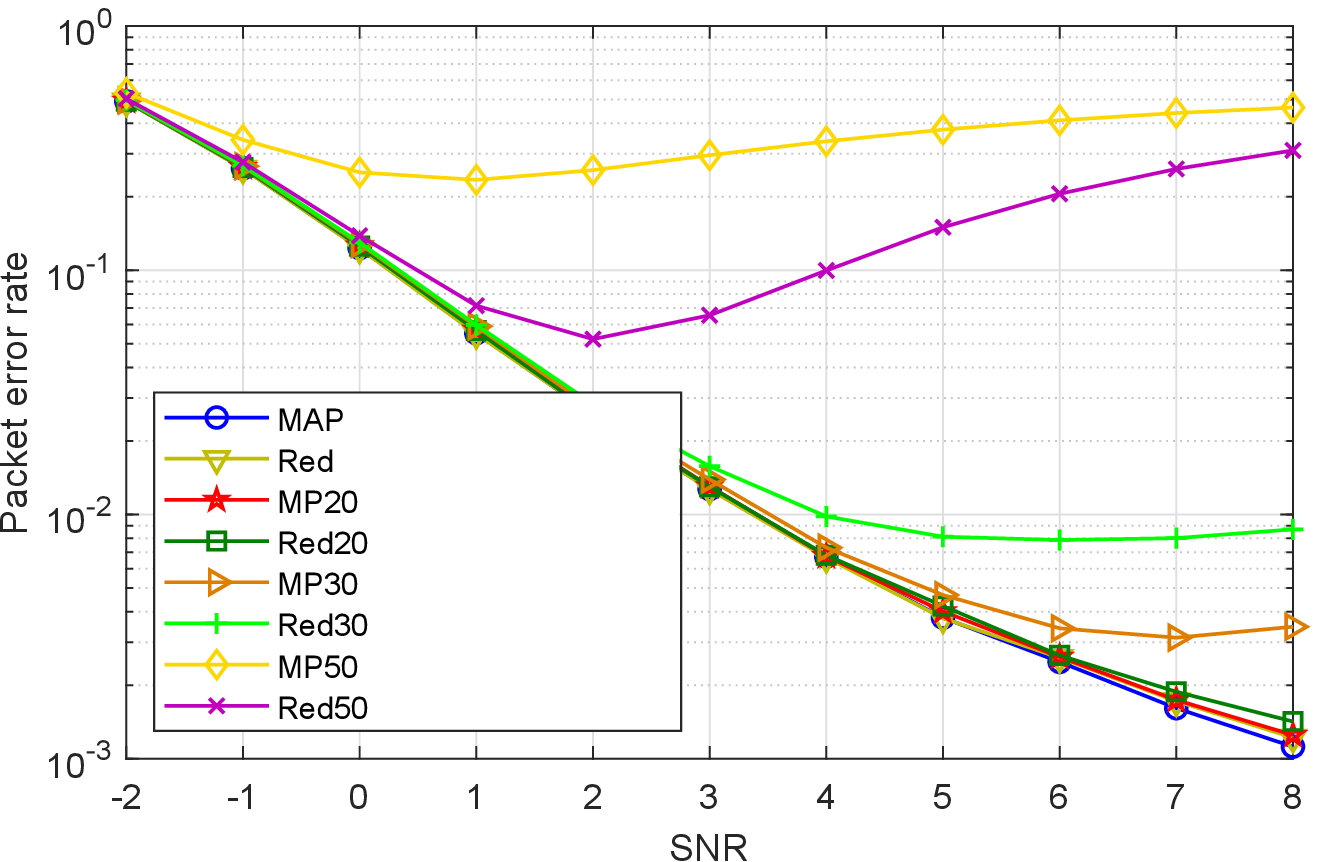}
\caption{Decoding performance of the full-state and reduced-state MAP detectors with imperfect estimation of the interference variance for a scenario with two interferer classes. $L_{I, 1}=2$, $L_{I, 2}=4$, $G_1=0.4$, $G_2=0.8$.}
\label{multiclass2}
\vspace*{-2mm}
\end{figure}
\else
\begin{figure}
\psfrag{SNR}{\hspace*{-5mm}\begin{small}$E_s/N_0$ [dB]\end{small}}
\psfrag{Packet error rate}{\begin{small}Packet error rate\end{small}}
\psfrag{MAP}{\begin{scriptsize}MAP, full, perfect\end{scriptsize}}
\psfrag{Red}{\begin{scriptsize}MAP, reduced, perfect\end{scriptsize}}
\psfrag{MP20}{\begin{scriptsize}MAP, full, 20$\%$ error\end{scriptsize}}
\psfrag{Red20}{\begin{scriptsize}MAP, reduced, 20$\%$ error\end{scriptsize}}
\psfrag{MP30}{\begin{scriptsize}MAP, full, 30$\%$ error\end{scriptsize}}
\psfrag{Red30}{\begin{scriptsize}MAP, reduced, 30$\%$ error\end{scriptsize}}
\psfrag{MP50}{\begin{scriptsize}MAP, full, 50$\%$ error\end{scriptsize}}
\psfrag{Red50}{\begin{scriptsize}MAP, reduced, 50$\%$ error\end{scriptsize}}
\includegraphics[width=0.49\textwidth]{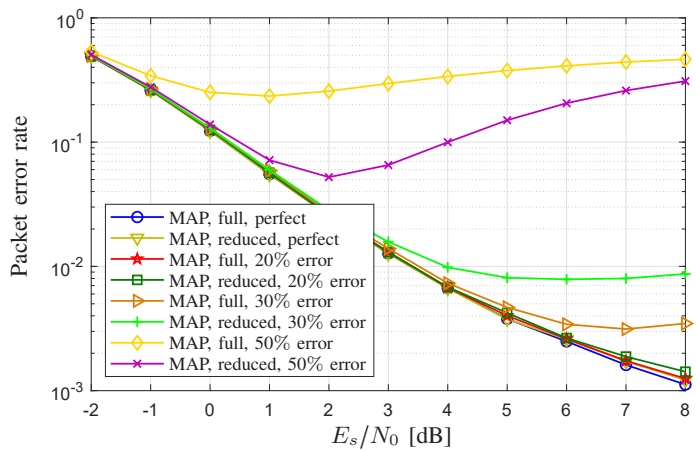}
\caption{Decoding performance of the full-state and reduced-state MAP detectors with imperfect estimation of the interference variance for a scenario with two interferer classes. $L_{I, 1}=2$, $L_{I, 2}=4$, $G_1=0.4$, $G_2=0.8$.}
\label{multiclass2}
\vspace*{-2mm}
\end{figure}
\fi
\subsection{Detection with scalable complexity}
In order to analyze the performance of the proposed method with scalable complexity according to Section \ref{sec:red}, we consider at first the achievable PER for different numbers of partitions. For this, we assume a simple scenario with two classes similar to Section \ref{sec:red}: $L_{I, 1}=2$, $L_{I, 2}=4$, $\sigma_{I, 1}^2=\sigma_{I, 2}^2=1$, $G_i=0.1\cdot L_{I, i}, \forall i$. The corresponding results for the achievable PER are depicted in Fig.\ \ref{multiclass3}.
\ifCLASSOPTIONdraftcls
\begin{figure}
\centering
\psfrag{SNR}{\hspace*{-5mm}\begin{normalsize}$E_s/N_0$ [dB]\end{normalsize}}
\psfrag{Packet error rate}{\begin{normalsize}Packet error rate\end{normalsize}}
\psfrag{MAP}{\begin{footnotesize}MAP, full\end{footnotesize}}
\psfrag{reduced}{\begin{footnotesize}MAP, reduced\end{footnotesize}}
\psfrag{scalable}{\begin{footnotesize}MAP, scalable, $P=2$\end{footnotesize}}
\psfrag{scalable P6}{\begin{footnotesize}MAP, scalable, $P=6$\end{footnotesize}}
\psfrag{scalable P10}{\begin{footnotesize}MAP, scalable, $P=10$\end{footnotesize}}
\includegraphics[width=0.7\textwidth]{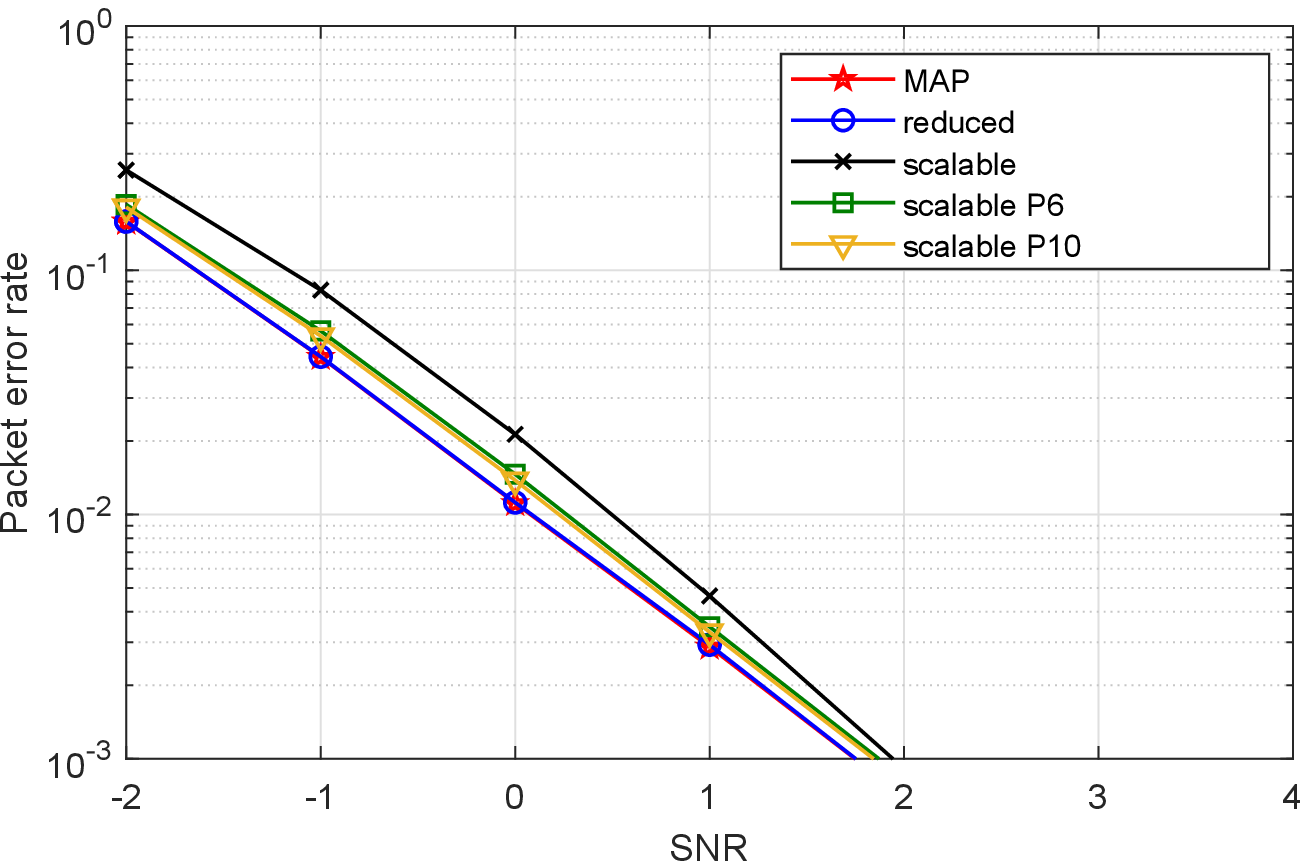}
\caption{Decoding performance of the full-state, reduced-state and scalable MAP detectors, respectively, for a scenario with two interferer classes. $L_{I, 1}=2$, $L_{I, 2}=4$, $\sigma_{I, 1}^2=\sigma_{I, 2}^2=1$, $G_i=0.1\cdot L_{I, i}, \forall i$.}
\label{multiclass3}
\vspace*{-2mm}
\end{figure}
\else
\begin{figure}
\psfrag{SNR}{\hspace*{-5mm}\begin{small}$E_s/N_0$ [dB]\end{small}}
\psfrag{Packet error rate}{\begin{small}Packet error rate\end{small}}
\psfrag{MAP}{\begin{scriptsize}MAP, full\end{scriptsize}}
\psfrag{reduced}{\begin{scriptsize}MAP, reduced\end{scriptsize}}
\psfrag{scalable}{\begin{scriptsize}MAP, scalable, $P=2$\end{scriptsize}}
\psfrag{scalable P6}{\begin{scriptsize}MAP, scalable, $P=6$\end{scriptsize}}
\psfrag{scalable P10}{\begin{scriptsize}MAP, scalable, $P=10$\end{scriptsize}}
\includegraphics[width=0.49\textwidth]{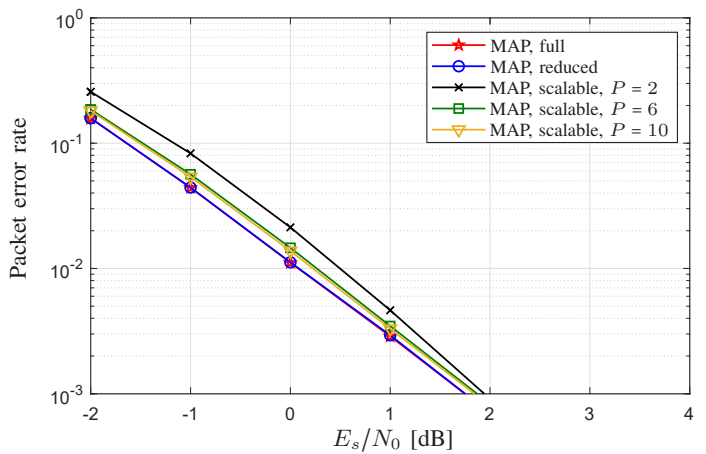}
\caption{Decoding performance of the full-state, reduced-state and scalable MAP detectors, respectively, for a scenario with two interferer classes. $L_{I, 1}=2$, $L_{I, 2}=4$, $\sigma_{I, 1}^2=\sigma_{I, 2}^2=1$, $G_i=0.1\cdot L_{I, i}, \forall i$.}
\label{multiclass3}
\vspace*{-2mm}
\end{figure}
\fi
Obviously, with a low number of partitions $P$ such as $P=2$, the quantization of the interference squared magnitudes is not sufficiently fine such that the PER is somewhat larger than the lower bound for the performance of any practical detector given by the performance of the full-state MAP detector. However, PER converges quickly to lower values with increasing $P$. In fact, the PERs obtained with $P=6$ and $P=10$ can hardly be distinguished from each other. We select $P=10$ for the following analysis. Also, we observe that the PER of the scalable detector is close to that obtained with the full-state MAP detector.  Hence, we conclude that this method is close-to-optimum.

For a real world scenario with a random distribution of interfering transmitters within a certain area around the receiver, we would obtain a continuous distribution of the interference variance which implies an infinite number of interference classes that need to be modeled via the Markov model for the optimum MAP detector and the reduced-state receiver according to Section \ref{sec:mult}. Among the solutions proposed in Section \ref{sec_3}, only the scalable detector is capable of providing a good performance at a moderate and adjustable complexity for this scenario. Hence, we compare the scalable detector only with the benchmark schemes for a scenario, where the nodes are randomly scattered (uniform distribution) within a ring-shaped area around the receiver with an inner radius of $100$ m and an outer radius of $1$ km. A path loss exponent of 3.5 and a carrier frequency of 869.5 MHz have been selected. Furthermore, each node is assigned to one of the three classes: $L_{I, 1}=10$, $L_{I, 1}=6$, or $L_{I, 1}=3$. We simulate two scenarios with different loads, where the load is equal for all interferer classes of the same scenario, i.e. $G_i=0.1,\forall i$ or $G_i=0.2,\forall i$. The corresponding results are depicted in Fig.\ \ref{multiclass4}.
\ifCLASSOPTIONdraftcls
\begin{figure}
\centering
\psfrag{SNR}{\hspace*{-5mm}\begin{normalsize}$E_s/N_0$ [dB]\end{normalsize}}
\psfrag{Packet error rate}{\begin{normalsize}Packet error rate\end{normalsize}}
\psfrag{genieLoad1}{\begin{footnotesize}genie-aided, $G_i=0.1$\end{footnotesize}}
\psfrag{genieLoad2}{\begin{footnotesize}genie-aided, $G_i=0.2$\end{footnotesize}}
\psfrag{erasureLoad1}{\begin{footnotesize}erasure, $G_i=0.1$\end{footnotesize}}
\psfrag{erasureLoad2}{\begin{footnotesize}erasure, $G_i=0.2$\end{footnotesize}}
\psfrag{const1}{\begin{footnotesize}const. var., $G_i=0.1$\end{footnotesize}}
\psfrag{const2}{\begin{footnotesize}const. var., $G_i=0.2$\end{footnotesize}}
\psfrag{scalable MAPLoad1}{\begin{footnotesize}MAP, scalable, $G_i=0.1$\end{footnotesize}}
\psfrag{scalable MAPLoad2}{\begin{footnotesize}MAP, scalable, $G_i=0.2$\end{footnotesize}}
\includegraphics[width=0.7\textwidth]{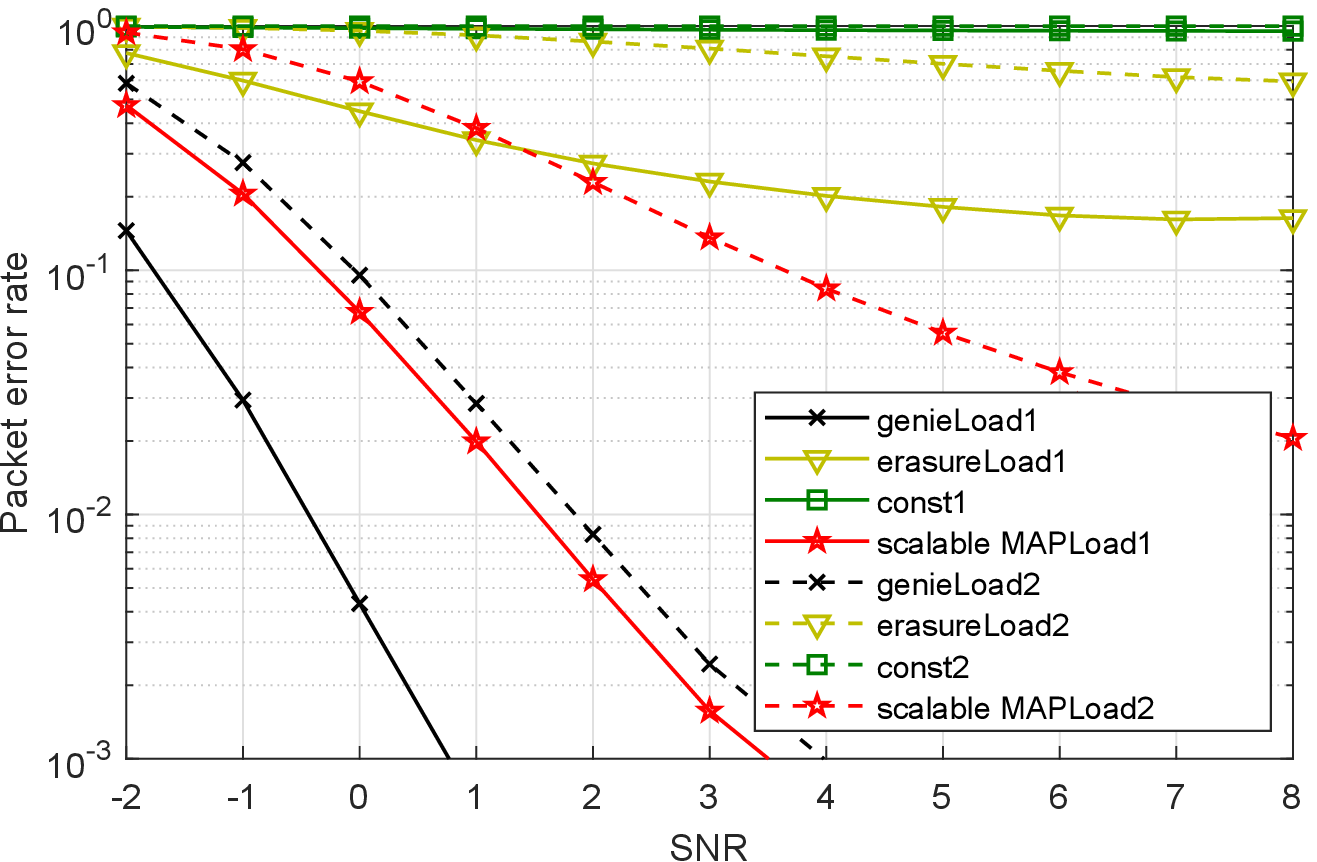}
\caption{Decoding performance of different detectors for a scenario with three interferer classes. $L_{I, 1}=10$, $L_{I, 2}=6$, $L_{I, 3}=3$, and $G_i=0.1,\forall i$ or $G_i=0.2,\forall i$. All transmitters are randomly distributed within a ring-shaped area around the receiver.}
\label{multiclass4}
\vspace*{-2mm}
\end{figure}
\else
\begin{figure}
\psfrag{SNR}{\hspace*{-5mm}\begin{small}$E_s/N_0$ [dB]\end{small}}
\psfrag{Packet error rate}{\begin{small}Packet error rate\end{small}}
\psfrag{genieLoad1}{\begin{scriptsize}genie-aided, $G_i=0.1$\end{scriptsize}}
\psfrag{genieLoad2}{\begin{scriptsize}genie-aided, $G_i=0.2$\end{scriptsize}}
\psfrag{erasureLoad1}{\begin{scriptsize}erasure, $G_i=0.1$\end{scriptsize}}
\psfrag{erasureLoad2}{\begin{scriptsize}erasure, $G_i=0.2$\end{scriptsize}}
\psfrag{scalable MAPLoad1}{\begin{scriptsize}MAP, scalable, $G_i=0.1$\end{scriptsize}}
\psfrag{scalable MAPLoad2}{\begin{scriptsize}MAP, scalable, $G_i=0.2$\end{scriptsize}}
\psfrag{const1}{\begin{scriptsize}const. var., $G_i=0.1$\end{scriptsize}}
\psfrag{const2}{\begin{scriptsize}const. var., $G_i=0.2$\end{scriptsize}}
\includegraphics[width=0.49\textwidth]{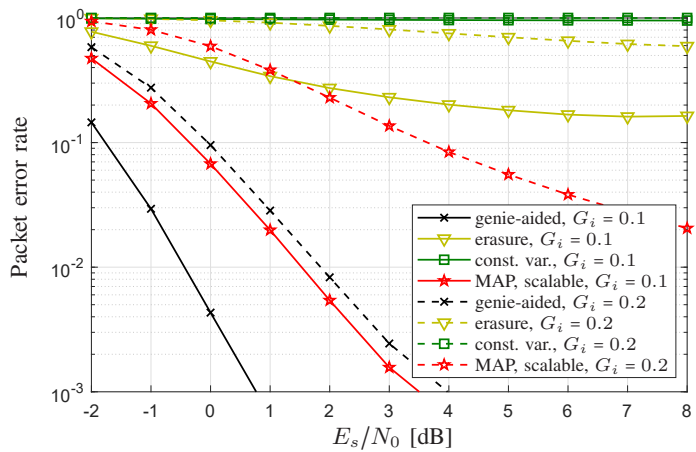}
\caption{Decoding performance of different detectors for a scenario with three interferer classes. $L_{I, 1}=10$, $L_{I, 2}=6$, $L_{I, 3}=3$, and $G_i=0.1,\forall i$ or $G_i=0.2,\forall i$. All transmitters are randomly distributed within a ring-shaped area around the receiver.}
\label{multiclass4}
\vspace*{-2mm}
\end{figure}
\fi
We observe a substantial performance degradation for all four methods, if the load increases from $G_i=0.1$ to $G_i=0.2$. 
Obviously, the proposed solution outperforms both the erasure method and the constant variance method. In particular, very low values of PER can be achieved using the scalable detector. For the erasure method, the PER does not reduce below $17\%$ even for high $E_s/N_0$ with $G_i=0.1$ and $60\%$ with $G_i=0.2$. For the constant variance detector, the PER remains close to '1', i.e. almost no error-free packet transmission is possible using this method in practice. The reason for this lies in the large fluctuations of the interference variance for the considered scenario. In fact, symbols corrupted by strong interference are typically interpreted as especially reliable by this detector, if the resulting receive signal magnitude is large. In addition, we observe a substantial gap between the PER achieved by the scalable detector and the PER of genie-aided detection, which is approx. 3 dB at $\mathrm{PER}=10^{-3}$ with $G_i=0.1$.

At last, we would like to investigate the system performance for this practical scenario under various code rates. We compare the PERs obtained for the code rates $R_c \in \{1/3, 2/5, 1/2\}$ depicted in Fig.\ \ref{multiclass5}, where the convolutional codes for the code rates $R_c=2/5$ and $R_c=1/2$, respectively, have been obtained by optimized puncturing of the code for $R_c=1/3$. We observe that the performance of all considered methods degrades noticeably with increasing code rate. For the genie-aided scheme, a degradation of 1.8 dB results when the code rate is increased from $R_c=1/3$ to $R_c=2/5$. With $R_c=1/2$, an error floor arises at high signal-to-noise ratios, such that a performance upper bound of $\mathrm{PER}\approx0.25\%$ is reached. For the erasure method, the performance degradation is less significant since the PER is already very high even for $R_c=1/3$. The proposed scalable MAP detector shows a degradation which is more substantial than for the genie-aided scheme at low code rates, i.e., up to 5 dB at a PER of $10^{-3}$ when comparing $R_c=1/3$ and $R_c=2/5$. Correspondingly, the performance gap between the scalable detector and the genie-aided scheme increases with increasing code rate, since inaccuracies in the interference state estimation and subsequent LLR calculation are compensated for less efficiently by the channel code. However, no practical method is known to date which can close this gap. Furthermore, we notice that the PER of the scalable MAP detector is very high ($\mathrm{PER}=3\%$ at $E_s/N_0=8$ dB) with $R_c=1/2$, which renders such a high code rate impractical for the considered scenario. In summary, for an LPWAN transmission in hostile environments with significant interference, a protection by sufficiently strong channel coding is recommended.
\ifCLASSOPTIONdraftcls
\begin{figure}
\centering
\psfrag{SNR}{\hspace*{-5mm}\begin{normalsize}$E_s/N_0$ [dB]\end{normalsize}}
\psfrag{Packet error rate}{\begin{normalsize}Packet error rate\end{normalsize}}
\psfrag{genieCode1}{\begin{footnotesize}genie-aided, $R_c=1/3$\end{footnotesize}}
\psfrag{genieCode2}{\begin{footnotesize}genie-aided, $R_c=2/5$\end{footnotesize}}
\psfrag{genieCode3}{\begin{footnotesize}genie-aided, $R_c=1/2$\end{footnotesize}}
\psfrag{erasureCode1}{\begin{footnotesize}erasure, $R_c=1/3$\end{footnotesize}}
\psfrag{erasureCode2}{\begin{footnotesize}erasure, $R_c=2/5$\end{footnotesize}}
\psfrag{erasureCode3}{\begin{footnotesize}erasure, $R_c=1/2$\end{footnotesize}}
\psfrag{scalableCode1}{\begin{footnotesize}MAP, scalable, $R_c=1/3$\end{footnotesize}}
\psfrag{scalableCode2}{\begin{footnotesize}MAP, scalable, $R_c=2/5$\end{footnotesize}}
\psfrag{scalableCode3}{\begin{footnotesize}MAP, scalable, $R_c=1/2$\end{footnotesize}}
\includegraphics[width=0.7\textwidth]{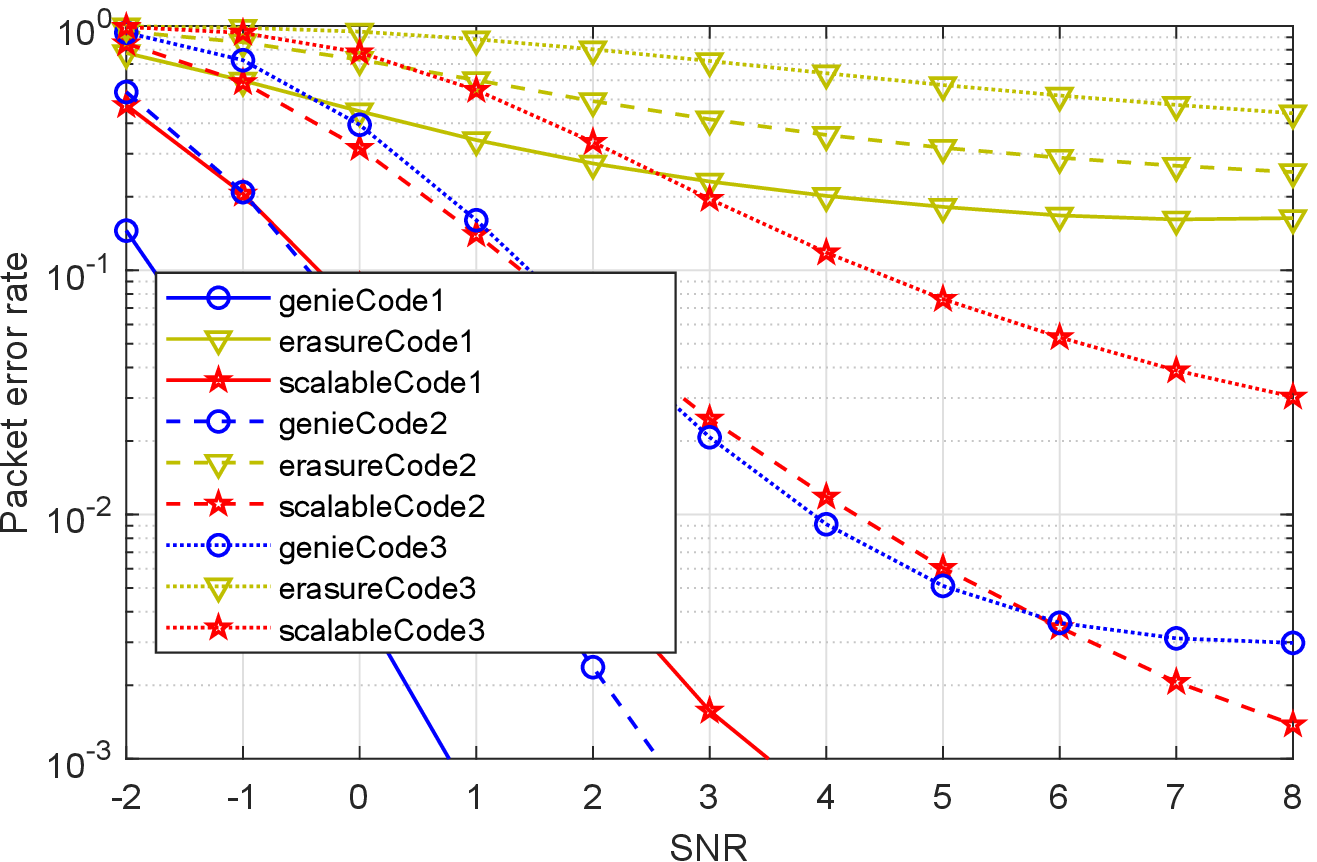}
\caption{Decoding performance of different detectors for a scenario with three interferer classes. $L_{I, 1}=10$, $L_{I, 2}=6$, $L_{I, 3}=3$, $G_i=0.1,\forall i$, and code rates $R_c=\{1/3, 2/5, 1/2\}$. All transmitters are randomly distributed within a ring-shaped area around the receiver.}
\label{multiclass5}
\vspace*{-2mm}
\end{figure}
\else
\begin{figure}
\psfrag{SNR}{\hspace*{-5mm}\begin{small}$E_s/N_0$ [dB]\end{small}}
\psfrag{Packet error rate}{\begin{small}Packet error rate\end{small}}
\psfrag{genieCode1}{\begin{scriptsize}genie-aided, $R_c=1/3$\end{scriptsize}}
\psfrag{genieCode2}{\begin{scriptsize}genie-aided, $R_c=2/5$\end{scriptsize}}
\psfrag{genieCode3}{\begin{scriptsize}genie-aided, $R_c=1/2$\end{scriptsize}}
\psfrag{erasureCode1}{\begin{scriptsize}erasure, $R_c=1/3$\end{scriptsize}}
\psfrag{erasureCode2}{\begin{scriptsize}erasure, $R_c=2/5$\end{scriptsize}}
\psfrag{erasureCode3}{\begin{scriptsize}erasure, $R_c=1/2$\end{scriptsize}}
\psfrag{scalableCode1}{\begin{scriptsize}MAP, scalable, $R_c=1/3$\end{scriptsize}}
\psfrag{scalableCode2}{\begin{scriptsize}MAP, scalable, $R_c=2/5$\end{scriptsize}}
\psfrag{scalableCode3}{\begin{scriptsize}MAP, scalable, $R_c=1/2$\end{scriptsize}}
\includegraphics[width=0.49\textwidth]{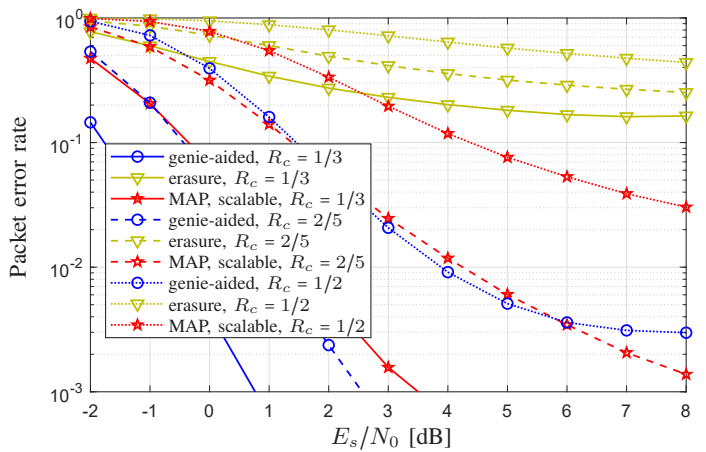}
\caption{Decoding performance of different detectors for a scenario with three interferer classes. $L_{I, 1}=10$, $L_{I, 2}=6$, $L_{I, 3}=3$, $G_i=0.1,\forall i$, and code rates $R_c=\{1/3, 2/5, 1/2\}$. All transmitters are randomly distributed within a ring-shaped area around the receiver.}
\label{multiclass5}
\vspace*{-2mm}
\end{figure}
\fi
\subsection{Complexity analysis}
Since some of the proposed methods aim at reaching close-to-optimum performance with reduced complexity, it is worth discussing the computational complexity of the proposed methods. Table \ref{tab:complexity} summarizes the complexity of the full MAP, reduced MAP and scalable MAP algorithms. Here, we took into account the worst-case complexity of the BCJR algorithm and did not consider possible suboptimal implementations, which typically rely on assumptions such as low probabilities in some branches which therefore can be discarded. The complexity of the BCJR algorithm can be given in terms of the number of states $S$ of the Markov product-chain per received symbol and the total required number of multiplications\footnote{Please note that in earlier work, the complexity analysis of the BCJR algorithm and its suboptimal versions has also included the memory requirements, cf.\ \cite{1055186}, \cite{4395266}. However, in recent times, memory has become less critical compared to the actual computational complexity. Hence, we provide the complexity with respect to the number of multiplications according to our implementation of the algorithms.}, which corresponds to $M=(L_{\mathrm{tot}}+2L_{\mathrm{add}})(1+S+3S^2)\approx 3(L_{\mathrm{tot}}+2L_{\mathrm{add}})S^2$. 
We also provide examples for the number of multiplications in some settings. For Example 1, we assume $L_{\mathrm{tot}}+2L_{\mathrm{add}}=36$, $K=2$, $L_{I, i}=5,\forall i$, and $P=3$. For Example 2, we select $L_{\mathrm{tot}}+2L_{\mathrm{add}}=36$, $K=2$, $L_{I, i}=10,\forall i$, and $P=5$. It can be observed from Table \ref{tab:complexity} that by employing the reduced MAP and scalable MAP algorithms, respectively, the computational complexity can be decreased by several orders of magnitude compared to the full MAP algorithm for the considered examples. Especially the scalable MAP algorithm appears well feasible in practical applications in terms of computational complexity.
\ifCLASSOPTIONdraftcls
\begin{table}[]
    \centering
    \caption{Complexity comparison of the three MAP methods.}
    \begin{tabular}{|l|c|c|c|c|}
    \hline
        \textbf{Method} & \textbf{Number of states} & \textbf{Number of multiplications} & \textbf{Example 1} & \textbf{Example 2}\\
        \hline
        \textbf{Full MAP} & $2\cdot2^{\sum_{i=1}^KL_{I, i}}$ & $12(L_{\mathrm{tot}}+2L_{\mathrm{add}})4^{\sum_{i=1}^KL_{I, i}}$ & $4.53\cdot10^8$ & $4.75\cdot10^{14}$\\
        \hline
        \textbf{Reduced MAP} & $2\prod_{i=1}^K(L_{I, i}+1)$ & $12(L_{\mathrm{tot}}+2L_{\mathrm{add}})\prod_{i=1}^K(L_{I, i}+1)^2$ & $5.6\cdot10^5$ & $6.3\cdot10^6$\\
        \hline
        \textbf{Scalable MAP} & $2P$ & $12(L_{\mathrm{tot}}+2L_{\mathrm{add}})P^2$ & $3.9\cdot10^3$ & $1.1\cdot10^4$\\
        \hline
    \end{tabular}
    \label{tab:complexity}
\end{table}
\else
\begin{table*}[]
    \centering
    \caption{Complexity comparison of the three MAP methods.}
    \begin{tabular}{|l|c|c|c|c|}
    \hline
        \textbf{Method} & \textbf{Number of states} & \textbf{Number of multiplications} & \textbf{Example 1} & \textbf{Example 2}\\
        \hline
        \textbf{Full MAP} & $2\cdot2^{\sum_{i=1}^KL_{I, i}}$ & $12(L_{\mathrm{tot}}+2L_{\mathrm{add}})4^{\sum_{i=1}^KL_{I, i}}$ & $4.53\cdot10^8$ & $4.75\cdot10^{14}$\\
        \hline
        \textbf{Reduced MAP} & $2\prod_{i=1}^K(L_{I, i}+1)$ & $12(L_{\mathrm{tot}}+2L_{\mathrm{add}})\prod_{i=1}^K(L_{I, i}+1)^2$ & $5.6\cdot10^5$ & $6.3\cdot10^6$\\
        \hline
        \textbf{Scalable MAP} & $2P$ & $12(L_{\mathrm{tot}}+2L_{\mathrm{add}})P^2$ & $3.9\cdot10^3$ & $1.1\cdot10^4$\\
        \hline
    \end{tabular}
    \label{tab:complexity}
\end{table*}
\fi
\section{Conclusion}
\label{sec_5}
In this paper, the optimal detector for LPWA networks in the presence of burst interference has been introduced. In the considered scenario, the telegram splitting method has been utilized in order to optimize the system performance. The amount of interference observed in each symbol interval has been modeled via a Markov process by exploiting the assumed knowledge of the interference arrival probability, the variance of a single interferer and the length of interferer packets. The resulting model is employed for the computation of a-posteriori symbol probabilities using a MAP detector from which LLR values can be determined for FEC decoding.
The proposed scheme outperforms a previously proposed erasure based detector noticeably.
Despite a substantial gap between the PER obtained using the proposed scheme and the PER of a genie-aided detector with perfect knowledge of the total interference variance, the proposed solution is capable of guaranteeing a high system performance. 

In addition, a reduced-state MAP detector has been proposed for scenarios with low-to-moderate interference load which performs very close to the full-state MAP detector. This solution entails a much lower computational complexity which enables the incorporation of multiple interferer classes such that the interference length and variance do not need to be identical for all interfering transmissions (as it was assumed for the full-state detector). However, a significant relaxation of the system assumptions allowing for the presence of a high number of classes and synchronization imperfections results in an increased complexity of the detector such that the number of interferer classes that can be supported remains still relatively low. Also, the considered full-state and reduced-state MAP detectors rely on the knowledge of the interference statistics, i.e., arrival probability, interference variance and packet length. In order to further reduce the complexity and eliminate the dependency of the detector on the knowledge of the interference statistics which may be not available in practice, we have proposed another detection method with scalable complexity. In this method, the number of states of the detector trellis diagram is a free parameter which can be selected by the system designer according to the tolerable complexity. This method shows a close-to-optimum performance (i.e., a performance close to that of the full-state MAP detector) even for a relatively small number of states of the underlying Markov model.
\section*{Appendix}
\subsection{Performance analysis of genie-aided detector}
In order to gain further insight into the performance of the proposed detectors under different interference conditions, we develop an analytical performance approximation to the genie-aided scheme which serves as an (approximate) upper bound to the performance of any realizable detector. For the genie-aided scheme, the signal-to-interference-and-noise ratio (SINR) varies within a codeword but is perfectly known for all positions of the codeword. In order to approximate the PER performance for a given channel code and given SINRs, we exploit an analogy to earlier work on cellular systems where link performance models have been designed for the application in system level simulations in order to 
determine the link PER at a reduced complexity with approximate expressions \cite{bruenighaus,brueck,nanda}.

In this context, it should be noted that in wireless systems
with varying fading and/or interference conditions and packet sizes which are not extremely large, the PER for a specific channel realization may be significantly different from the average PER, and a carefully designed link level performance model is needed which accounts for the specifics of the given channel snapshot. In the literature, various such models have been developed for the usage in system level simulations, 
cf.\ e.g.\ \cite{bruenighaus,brueck,nanda}. Typically, in such models a set of quality measures is collected for the different positions within the codeword such as raw bit error rates, SINRs and channel capacities according to the current channel realization. 
Subsequently, a single quality measure is derived from the available set of measures which is used to derive the anticipated PER for the current channel conditions. For the following, we adopt a specific approach described in 
\cite{bruenighaus,brueck} where an {\em effective} SINR is used for a quick estimation of PER, assuming random interleaving within a codeword. 
The effective SINR approach has been justified in \cite{nanda} by
considering the high probability error events in optimum soft-decision decoding of a convolutional code via the Viterbi algorithm.
In general, the effective SINR for random interleaving is expressed as \cite{bruenighaus}[Eq.\ (1)]
\begin{equation}
{\rm SINR}_{\rm eff} = b \, I^{-1} \left( \frac{1}{n} \, \sum\limits_{l=0}^{n-1} 
I \left( \frac{{\rm SINR}_l}{b} \right) \right),
\label{mapping_eq}
\end{equation}
where $I(\cdot)$ and $I^{-1}(\cdot)$ are a function depending on the employed model and its inverse, respectively, ${\rm SINR}_l$ denotes the SINR experienced by the codeword symbol $c[l]$, and $b$ is a constant which can be used for optimization of the PER estimation.

In the next step, the PER is estimated from the
effective SINR by applying a nonlinear function. In \cite{bruenighaus,brueck}, it is proposed to use either the PER vs.\ SNR characteristics of the adopted convolutional code for the AWGN channel or an optimized spline curve.
While the latter approach allows for a higher estimation accuracy, 
the former one is more intuitive.

For a detailed discussion of possible choices for the function $I(\cdot)$ we refer to \cite{bruenighaus}. Since our main priority lies on intuitive expressions instead of maximum estimation accuracy we  select the {\em capacity approach} corresponding to
$I(y) = {\rm ld}(1+y)$,
which results in a performance model referred to as capacity
effective SINR metric (CESM) \cite{bruenighaus}.

In the following, we assume that the SINRs of all codeword positions are moderate-to-high which allows for further simplifications. Since  $I(y) \approx {\rm ld}(y)$ and $I^{-1}(y)\approx 2^y$ in this regime, the effective SINR in (\ref{mapping_eq}) can be approximately expressed as
\begin{equation}
  {\rm SINR}_{\rm eff} \approx \sqrt[n]{ \prod\limits_{l=0}^{n-1} {\rm SINR}_l}.
  \label{eq_SINR_eff_1}
  \end{equation}
Accordingly, the effective SINR can be expressed via the {\em geometric mean} of the individual SINRs of the codeword positions.

For a performance estimation of the considered genie-aided detector under given interference conditions, the required set of SINRs within a codeword is determined as
\begin{equation}
  {\rm SINR}_{l} = \frac{1}{\sigma^2_{z,i}[m]+\sigma_N^2},
  \label{eq_SINR_eff_2}
  \end{equation}
where $\sigma^2_{z,i}[m]$ denotes the total interference variance for symbol $x_i[m]$ corresponding to codeword symbol $c[l]$.
Finally, the PER of the adopted convolutional code under the given interference conditions is estimated  by the PER of that code for an AWGN channel with ${\rm SNR}={\rm SINR}_{\rm eff}$.

For randomly changing sets of SINRs for different codewords, an approximate performance bound can be obtained by averaging the PER estimate with respect to the interference statistics.
\bibliographystyle{IEEEtran}
\bibliography{Literature}

\begin{thebibliography}{10}
\providecommand{\url}[1]{#1}
\csname url@samestyle\endcsname
\providecommand{\newblock}{\relax}
\providecommand{\bibinfo}[2]{#2}
\providecommand{\BIBentrySTDinterwordspacing}{\spaceskip=0pt\relax}
\providecommand{\BIBentryALTinterwordstretchfactor}{4}
\providecommand{\BIBentryALTinterwordspacing}{\spaceskip=\fontdimen2\font plus
\BIBentryALTinterwordstretchfactor\fontdimen3\font minus
  \fontdimen4\font\relax}
\providecommand{\BIBforeignlanguage}[2]{{%
\expandafter\ifx\csname l@#1\endcsname\relax
\typeout{** WARNING: IEEEtran.bst: No hyphenation pattern has been}%
\typeout{** loaded for the language `#1'. Using the pattern for}%
\typeout{** the default language instead.}%
\else
\language=\csname l@#1\endcsname
\fi
#2}}
\providecommand{\BIBdecl}{\relax}
\BIBdecl

\bibitem{own}
S.~Kisseleff, J.~Kneissl, G.~Kilian, and W.~Gerstacker, ``{Optimal {MAP}
  Detection in Presence of Burst Interference for Low Power Wide Area
  Networks},'' in \emph{Proc. of IEEE Global Communications Conference}, Dec.
  2018, pp. 1--6.

\bibitem{7815384}
U.~Raza, P.~Kulkarni, and M.~Sooriyabandara, ``{Low Power Wide Area Networks:
  An Overview},'' \emph{IEEE Communications Surveys $\&$ Tutorials}, vol.~19,
  no.~2, pp. 855--873, 2017.

\bibitem{8030482}
F.~{Adelantado}, X.~{Vilajosana}, P.~{Tuset-Peiro}, B.~{Martinez},
  J.~{Melia-Segui}, and T.~{Watteyne}, ``{Understanding the Limits of
  LoRaWAN},'' \emph{IEEE Communications Magazine}, vol.~55, no.~9, pp. 34--40,
  Sept. 2017.

\bibitem{7721743}
M.~Centenaro, L.~Vangelista, A.~Zanella, and M.~Zorzi, ``{Long-range
  communications in unlicensed bands: the rising stars in the IoT and smart
  city scenarios},'' \emph{IEEE Wireless Communications}, vol.~23, no.~5, pp.
  60--67, Oct. 2016.

\bibitem{7132717}
E.~Ndih, S.~Cherkaoui, and I.~Dayoub, ``{Analytic Modeling of the Coexistence
  of IEEE 802.15.4 and IEEE 802.11 in Saturation Conditions},'' \emph{IEEE
  Communications Letters}, vol.~19, no.~11, pp. 1981--1984, Nov. 2015.

\bibitem{8422801}
J.~Robert, S.~Rauh, H.~Lieske, and A.~Heuberger, ``{IEEE 802.15 Low Power Wide
  Area Network (LPWAN) PHY Interference Model},'' in \emph{Proc. of IEEE
  International Conference on Communications}, May 2018, pp. 1--6.

\bibitem{7575656}
H.~Lieske, G.~Kilian, M.~Breiling, S.~Rauh, J.~Robert, and A.~Heuberger,
  ``{Decoding Performance in Low-Power Wide Area Networks With Packet
  Collisions},'' \emph{IEEE Trans. on Wireless Commun.}, vol.~15, no.~12, pp.
  8195--8208, Dec. 2016.

\bibitem{6525243}
G.~Kilian, H.~Petkov, R.~Psiuk, H.~Lieske, F.~Beer, J.~Robert, and
  A.~Heuberger, ``{Improved Coverage for Low-Power Telemetry Systems using
  Telegram Splitting},'' in \emph{Proc. of European Conference on Smart
  Objects, Systems and Technologies (Smart SysTech)}, June 2013.

\bibitem{6999938}
G.~Kilian, M.~Breiling, H.~Petkov, H.~Lieske, F.~Beer, J.~Robert, and
  A.~Heuberger, ``{Increasing Transmission Reliability for Telemetry Systems
  Using Telegram Splitting},'' \emph{IEEE Trans. on Commun.}, vol.~63, no.~3,
  pp. 949--961, March 2015.

\bibitem{6547827}
J.~Lin, M.~Nassar, and B.~Evans, ``Impulsive noise mitigation in powerline
  communications using sparse {Bayesian} learning,'' \emph{IEEE Journal on
  Selected Areas in Communications}, vol.~31, no.~7, pp. 1172--1183, July 2013.

\bibitem{990732}
M.~Zimmermann and K.~Dostert, ``Analysis and modeling of impulsive noise in
  broad-band powerline communications,'' \emph{IEEE Trans. on Electromagnetic
  Compatibility}, vol.~44, no.~1, pp. 249--258, 2002.

\bibitem{fertonani2009reliable}
D.~Fertonani and G.~Colavolpe, ``On reliable communications over channels
  impaired by bursty impulse noise,'' \emph{IEEE Trans. on Communications},
  vol.~57, no.~7, 2009.

\bibitem{5504595}
J.~Mitra and L.~Lampe, ``Convolutionally coded transmission over
  {Markov-Gaussian} channels: Analysis and decoding metrics,'' \emph{IEEE
  Trans. on Communications}, vol.~58, no.~7, pp. 1939--1949, July 2010.

\bibitem{6824847}
W.~Zhang, M.~Suresh, R.~Stoleru, and H.~Chenji, ``On modeling the coexistence
  of 802.11 and 802.15.4 networks for performance tuning,'' \emph{IEEE Trans.
  on Wireless Communications}, vol.~13, no.~10, pp. 5855--5866, Oct. 2014.

\bibitem{proakis}
J.~Proakis, \emph{{Digital Communications}}, 4th~ed.\hskip 1em plus 0.5em minus
  0.4em\relax McGraw-Hill, New York, 2001.

\bibitem{markov}
W.~Stewart, \emph{{Probability, Markov Chains, Queues, and Simulation: The
  Mathematical Basis of Performance Modeling}}.\hskip 1em plus 0.5em minus
  0.4em\relax Princeton University Press, NJ, 2009.

\bibitem{1055186}
L.~Bahl, J.~Cocke, F.~Jelinek, and J.~Raviv, ``Optimal decoding of linear codes
  for minimizing symbol error rate,'' \emph{IEEE Trans. on Information Theory},
  vol.~20, no.~2, pp. 284--287, March 1974.

\bibitem{dempster1977maximum}
A.~Dempster, N.~Laird, and D.~Rubin, ``{Maximum likelihood from incomplete data
  via the EM algorithm},'' \emph{Journal of the Royal Statistical Society.
  Series B (Methodological)}, pp. 1--38, 1977.

\bibitem{1056489}
S.~Lloyd, ``Least squares quantization in {PCM},'' \emph{IEEE Trans. on
  Information Theory}, vol.~28, no.~2, pp. 129--137, March 1982.

\bibitem{7564982}
F.~Sacuto, G.~Ndo, F.~Labeau, and B.~Agba, ``{MAP} optimum receiver mitigating
  correlated impulsive noise,'' in \emph{Proc. of IEEE Wireless Communications
  and Networking Conference}, April 2016, pp. 1--6.

\bibitem{4395266}
D.~{Fertonani}, A.~{Barbieri}, and G.~{Colavolpe}, ``Reduced-complexity {BCJR}
  algorithm for turbo equalization,'' \emph{IEEE Transactions on
  Communications}, vol.~55, no.~12, pp. 2279--2287, 2007.

\bibitem{bruenighaus}
K.~Brueninghaus, D.~Astely, T.~Salzer, S.~Visuri, A.~Alexiou, S.~Karger, and
  G.-A. Seraji, ``Link performance models for system level simulations of
  broadband radio access systems,'' in \emph{IEEE International Symposium on
  Personal, Indoor and Mobile Radio Communications}, vol.~4, 2005, pp.
  2306--2311.

\bibitem{brueck}
S.~{Bruck}, ``Modeling interference diversity in {GSM} networks,'' in
  \emph{Vehicular Technology Conference Fall}, vol.~1, 2000, pp. 459--466.

\bibitem{nanda}
S.~Nanda and K.~Rege, ``Frame error rates for convolutional codes on fading
  channels and the concept of effective {$E_b/N_0$},'' in \emph{Proceedings of
  GLOBECOM}, 1995, pp. 27--32.

\end{thebibliography}
\end{document}